\documentclass{amsart}
\usepackage{amsmath}
\usepackage{amsthm}
\usepackage{amsfonts}
\usepackage{amssymb}
\usepackage{amsbsy}
\usepackage{amscd}
\usepackage{eucal}
\usepackage[dvips]{graphicx}
\usepackage{amsaddr}

\newtheorem{theorem}{Theorem}
\newtheorem{lemma}[theorem]{Lemma}
\newtheorem{proposition}[theorem]{Proposition}

\theoremstyle{definition}

\numberwithin{equation}{section}

\newcommand{\beq}{\begin{equation}}
\newcommand{\eeq}{\end{equation}}

\def\staccrel#1#2{\mathrel{\mathop{#1}\limits_{#2}}}

\begin{document}

\title[Geometrical theory of diffracted rays, orbiting, and complex rays]{Geometrical theory of diffracted rays, orbiting and complex rays}

\author[E. De Micheli]{Enrico De Micheli}
\address{\sl IBF -- Consiglio Nazionale delle Ricerche \\ Via De Marini, 6 - 16149 Genova, Italy \\
E-mail: enrico.demicheli@cnr.it}
\author[G. A. Viano]{Giovanni Alberto Viano}
\address{\sl Dipartimento di Fisica -- Universit\`a di Genova,\\
Istituto Nazionale di Fisica Nucleare -- Sezione di Genova, \\
Via Dodecaneso, 33 - 16146 Genova, Italy \\
E--mail: viano@ge.infn.it}


\begin{abstract}
In this article, the ray tracing method is studied beyond the classical geometrical
theory. The trajectories are here regarded as geodesics in a Riemannian manifold,
whose metric and topological properties are those induced by the refractive
index (or, equivalently, by the potential). First, we derive the geometrical
quantization rule, which is relevant to describe the orbiting bound--states
observed in molecular physics. Next, we derive properties of the diffractive
rays, regarded here as geodesics in a Riemannian manifold with boundary.
A particular attention is devoted to the following problems:
(i) modification of the classical stationary phase method suited to a
neighborhood of a caustic;
(ii) derivation of the connection formulae which enable one to obtain the uniformization
of the classical eikonal approximation by patching up geodesic segments
crossing the axial caustic;
(iii) extension of the eikonal equation to mixed hyperbolic--elliptic systems,
and generation of complex--valued rays in the shadow of the caustic.
By these methods, we can study the creeping waves in
diffractive scattering, describe the orbiting resonances present in molecular
scattering beside the orbiting bound--states, and, finally, describe the
generation of the evanescent waves, which are relevant in the nuclear rainbow.
\end{abstract}

\maketitle

\noindent
{\it ``Lumen propagatur seu diffunditur non solum directe, refracte ac reflexe,
sed etiam alio quodam quarto modo, diffracte.''}
\begin{flushright}
Francesco Maria Grimaldi \\
(Physico--Mathesis de lumine, coloribus, et iride, aliisque adnexis libri duo;
Bononiae 1665)
\end{flushright}

\section{Introduction}
\label{se:introduction}

In the decade 1950--1960, J. B. Keller \cite{Keller1,Levy,Hansen} wrote several
papers in which he introduced the so--called {\it Geometrical Theory of Diffraction}
(GTD). This theory can be regarded as an extension of geometrical optics, which
accounts for diffraction by introducing the diffracted rays in addition to the
usual rays of geometrical optics. Nowadays, GTD includes features which were not
present in Keller's first derivation such as, for instance, uniform solutions at
a caustic. It was, in fact, Ludwig \cite{Ludwig1} who obtained a uniform
asymptotic expansion of the wavefield at the caustic, which is the locus where
the rays of geometrical optics have an envelope and the amplitude presents a
singularity. Ludwig derived a formal asymptotic series, which is uniformly
valid. After these seminal works, there has been a steady flow of papers
addressing various aspects of the theory. On one hand, papers oriented to pure
and applied electromagnetic theory, like radiation and scattering of waves,
antenna design, waveguide theory, and so on \cite{Hansen}; on the other hand,
a highly theoretical and mathematically sophisticated theory of propagation
of singularities, and diffraction of waves on manifolds \cite{Melrose}.
Moreover, in theoretical physics, GTD is closely related to the Feynman integral
approach. It is worth mentioning the recent paper by Schaden and Spruch
\cite{Schaden}, where a link between diffraction in the
semiclassical approximation and Feynman's path integral representation of
the Green function can be found. These authors note that the evaluation of the Casimir
effects needs the consideration of diffraction, at least semiclassically.
Along with Casimir effects, the calculation of the exchange
second virial coefficient of a hard--sphere gas evaluated by E. H. Lieb by means
of the path integral should be recalled \cite{Lieb}. This problem is indeed similar
to the classical problem of diffraction of waves (of short wavelength)
around a sphere into the dark zone \cite{Lieb}. Last but not least, GTD
also suggested the introduction of the complex angular momenta (CAM) in
wave mechanics \cite{DeAlfaro}, which is not surprising since the most
precise diffraction measurements can be performed in nuclear and particle
physics.

A particularly useful mathematical tool in diffraction theory is the Watson
resummation of the partial wave expansion. Unfortunately, working with this
transform leads quite frequently to technical convergence difficulties; this is
the case, for instance, in the computation of the virial coefficient
\cite{Lieb}, and in the CAM theory of potential scattering if one considers
potentials far beyond the Yukawian class. In this situation, one is naturally
led to leave out the Watson transform and to push further the path integral
approach. However, also in this case, a price has to be paid since one is forced to
introduce various conjectures based on physical intuition. As a typical example,
we can mention the splitting of grazing rays at the boundary of the
obstacle and, accordingly, the generation of the diffracted rays. At this point
a question emerges:

~

{\bf Problem:} Is it possible to prove the conjectures of GTD by using the
methods supplied by the differential geometry?

~

We believe that, at the moment, it is very hard to give a general and definitive
positive answer. Nevertheless, an answer in some specific physical problem can
be found. The purpose of this paper is precisely that of providing a geometrical
theory of the following physical phenomena: \\
\indent (a) orbiting resonances and bound states, which are observed in molecular
scattering \cite{Toennies,DeMicheli2}; \\
\indent (b) diffractive scattering at high--energy in nuclear and particle physics
\cite{Hagiwara,Lennox,DeMicheli4};\\
\indent (c) the generation of evanescent waves in the shadow of the caustic, which are
relevant for the nuclear rainbow \cite{Delbar,Brink}.

In a previous paper \cite{DeMicheli1} we have obtained very preliminary
results concerning problem (b): i.e., the diffractive scattering at high
energy by a compact, convex and opaque object. Nevertheless, in spite of these
efforts, many relevant problems remained. It is precisely the main purpose
of the present article to fill this gap and, more specifically, to study in detail
the following problems: \\
1) A method of geometric quantization based on a procedure of
analytic continuation, appropriate for describing the spectrum of bound states in
molecular physics. This problem will be addressed in Section \ref{se:geometrical}. \\
2) The geometrical structure of the caustic in diffractive scattering,
considering separately the axial caustic from the caustic generated by the border of
the diffracting obstacle. This question will be treated in Section \ref{se:riemannian}. \\
3) Still in Section \ref{se:riemannian} we derive eikonal equations of
Ludwig--type, by using the Chester-Friedman-Ursell \cite{Chester} approximation,
which is a modification of the stationary phase method, suitable in the neighborhood
of a caustic. In the same section, the mixed (hyperbolic--elliptic) character
of the Ludwig system is proved and analyzed. \\
4) Section \ref{se:singular} is divided in three parts. In the first
one, we derive the scattering amplitude at high energy of the creeping waves
in diffractive theory by making use of an appropriate semi--phenomenological
transport equation. In particular, we obtain an expression of the damping
factor of the creeping waves, which depends upon the curvature of the obstacle,
and it is related to the breaking of quantization. The second part is
devoted to the analysis of the orbiting resonances, which are present in
molecular physics, and some explicit expressions of the phase--shifts
are obtained. In the third part of Section \ref{se:singular}, we prove the existence
of the evanescent waves in the shadow of the caustic, which follows from the
elliptic character of the Ludwig eikonal system. \\
5) Finally, some conclusions are drawn in Section \ref{se:conclusions}.

\section{Geometrical theory of orbiting bound states}
\label{se:geometrical}
\subsection{Geometric preliminaries}
\label{subse:physical}
Let ${\mathbb{M}}^n$ denote a Riemannian manifold, let ${\mathcal{C}}$ be a chart
of ${\mathbb{M}}^n$, let ${\mathbf{x}}=(x_1,\ldots,x_n)$ be the local coordinates
in ${\mathcal{C}}$, and let $g_{ij}$ be the metric tensor. As usual
$g = \left|\det (g_{ij})\right|$, and $g^{ij}=(g_{ij})^{-1}$. Consider
the Helmholtz equation
\begin{equation}
\Delta\psi + k^2\psi=0,
\label{2.1}
\end{equation}
where $k$ is the wavenumber and $\Delta$ is the Laplace--Beltrami operator which,
when applied to a function $\psi \in C^\infty({\mathbb{M}}^n)$, reads as
\begin{equation}
\Delta \psi = \frac{1}{\sqrt g} \, \sum_{i=1}^n \, \frac{\partial}{\partial x_i}
\left(\sum_{j=1}^n g^{ij}\,\sqrt g\,\frac{\partial}{\partial x_j}\right) \psi.
\label{2.2}
\end{equation}
We now introduce the infinite--dimensional space $\Omega({\mathbb{M}}^n; p,q)$
of the piecewise differentiable paths $c$ connecting the
point $p$ with the point $q$ on ${\mathbb{M}}^n$: i.e., if $p$ and $q$ are two fixed points on
${\mathbb{M}}^n$, then $c:[0,1]\rightarrow{\mathbb{M}}^n$ is a piecewise
differentiable path such that $c(0)=p$ and $c(1)=q$. To any element
$c \in \Omega({\mathbb{M}}^n; p,q)$ we associate an infinite--dimensional
vector space $T_c\Omega$, which can be identified with the space tangent to
$\Omega$ at the point $c$. More precisely, $T_c\Omega$ is the vector space
formed by all fields of piecewise differentiable vectors $V$ along the
path $c$ such that $V(0)=0$ and $V(1)=0$. Next, we consider the action $S$ as
a functional on the paths $\{c\}$ \cite{Doubrovine}:
\begin{equation}
S(c) = \int_0^1\left|\frac{dc}{dt}\right|^2\,dt=
\int_0^1 g_{ij}\,\dot{x}_i \, \dot{x}_j\,dt.
\label{2.3}
\end{equation}
From the first variation of the functional $S$ we can derive that the extremals
of the functional $S(c)$ are represented by geodesics $c(t)=\gamma(t)$
parametrized by $t$. Let us now consider the second variation $S_{**}$ of the
functional $S$ along the geodesic $\gamma$; it is referred to as the {\it Hessian}
of $S$. For a given geodesic $\gamma=\gamma(t)$, $0\leqslant t \leqslant 1$, a
geodesic variation of $\gamma$ is a one--parameter family of geodesics
$\gamma_s=\gamma_s(t)$ $(-\epsilon\leqslant s \leqslant\epsilon)$ such that
$\gamma_0=\gamma$. For each fixed $s$, $\gamma_s(t)$ describes a geodesic as
$t$ varies from $0$ to $1$. Each variation gives rise to an infinitesimal
variation, which is a vector field defined along $\gamma$. The points
$\gamma(0)$ and $\gamma(1)$ of $\gamma$ are said to be \emph{conjugate} if there is a
variation $\gamma_s$ inducing an infinitesimal variation vanishing at
$t=0$ and $t=1$. We then denote by $\mu$ the index of the Hessian, i.e., the
largest dimension of the subspace $T_\gamma\Omega$ on which $S_{**}$ is negative
definite. Then the following classical theorem due to Morse \cite{Milnor2} can
be stated.
\begin{proposition}{{\rm(Morse Index Theorem \cite{Milnor2})}}
\label{pro:1}
The index $\mu$ of the hessian $S_{**}$ is equal to the number of points belonging
to $\gamma(t)$ which are conjugate to the initial point $p = \gamma(0)$, counted according to 
their multiplicity.
\end{proposition}
In the theory of the orbiting bound states, the manifold ${\mathbb{M}}^n$ under
consideration is ${\mathbb{M}}^n={\mathbb{S}}^2$. The unit sphere can be described
by the angular coordinates $\theta$ and $\varphi$; accordingly, the matrix
elements $g_{ij}$ have the values: $g_{11}=1$, $g_{22}=\sin^2\theta$,
$g_{12}=g_{21}=0$. Now, choose a point $p_0\in{\mathbb{S}}^2$; all points of
${\mathbb{S}}^2$, except for the antipodal one (say $q_0$), are connected to $p_0$
by only one geodesic of minimal length, whereas the antipodal point $q_0$ is
connected to $p_0$ by a continuum of geodesics of minimal length, which can be
obtained as the intersections of ${\mathbb{S}}^2$ with the planes passing
through the diameter of ${\mathbb{S}}^2$ connecting $p_0$ to $q_0$. By rotating
these planes, and keeping $p_0$ and $q_0$ fixed, we obtain a variation vector
field which is a Jacobi field \cite{Milnor2} vanishing at $p_0$ and $q_0$.
Thus, we can conclude that $p_0$ and $q_0$ are conjugate with multiplicity one
because the possible rotations are along one direction only. Then, in view of
the Morse Index Theorem, we can claim that the index $\mu$ of $S_{**}$ jumps
by one when the geodesic $\gamma(t)$, whose initial point is $\gamma(0)=p_0$,
crosses $q_0$.

Now, we introduce the momenta associated with the coordinates $\theta$,
$\varphi$ through the Legendre transformation,
\begin{subequations}
\label{2.4}
\begin{eqnarray}
p_\theta &=& \frac{\partial S}{\partial\theta}, \label{2.4a} \\
p_\varphi &=& \frac{\partial S}{\partial\varphi}, \label{2.4b}
\end{eqnarray}
\end{subequations}
whenever these equations yield a diffeomorphism of some sufficiently small
neighborhoods $U_0$, $V_0$ of the points $(\theta^0,\varphi^0)$,
$(p_\theta^0, p_\varphi^0)$, respectively. We are thus led to the phase space
described by coordinates and momenta $(\theta,\varphi;p_\theta,p_\varphi)$.
In the phase space, one generates a family of trajectories, which forms a manifold
$\Lambda$ smoothly embedded in the phase space. All the trajectories in this
family have the same energy; this condition can be stated by setting the total
Hamiltonian $g^{ij}p_i p_j=H={\mbox{Const.}}$ (without loss of generality,
the constant can be set equal to $1$). Next, the first integrals $h$ of the
Hamiltonian system described by $H$ can be determined by equating the
Poisson brackets to zero, i.e., $\{h,H\}=0$. We can thus obtain \cite{Maslov}
\begin{subequations}
\label{2.5}
\begin{eqnarray}
&&p_\varphi^2 = c_1^2 = {\mbox{Const.}}, \label{2.5a} \\
&&p_\theta^2 + \frac{c_1^2}{\sin^2\theta} = 1, \label{2.5b}
\end{eqnarray}
\end{subequations}
Equation (\ref{2.5b}) defines a smooth curve $\gamma$ in the domain
$0<\theta<\pi$, $p_\theta \in{\mathbb{R}}$, which is diffeomorphic to the circle.
Equations (\ref{2.5}) determine a family of invariant Lagrangian
manifold $\Lambda$ diffeomorphic to the torus $T$ and smoothly depending on $c_1$.

\subsection{Oscillating integrals and eikonal approximation}
\label{subse:oscillating}
We now look for a solution to (\ref{2.1}) of the following form:
\begin{equation}
\psi({\mathbf{x}};k)=
\int A({\mathbf{x}};s)\,e^{{\rm i} k \Phi({\mathbf{x}};s)}\,ds,
\label{2.6}
\end{equation}
where $\Phi({\mathbf{x}};s)$ is the Hamilton characteristic function
\cite{Goldstein}, and $s$ is a coordinate on the path
connecting the source located at $p_0$ with the point of coordinate
${\mathbf{x}}$. The r.h.s. of (\ref{2.6}) is an integral of oscillating type.
The principal contribution to $\psi({\mathbf{x}},k)$, as $k\rightarrow\infty$,
corresponds to the stationary point of $\Phi$, in the neighborhood of which the
exponential ceases to oscillate rapidly. These stationary points can be obtained
from the equation $\frac{\partial \Phi}{\partial s} = 0$ (provided that
$\frac{\partial^2 \Phi}{\partial s^2} \neq 0$): here for $s$ we take the
arc length coordinate of the paths. If condition
$\frac{\partial \Phi}{\partial s} = 0$ is satisfied by a unique value $s_0$ of
$s$, which corresponds to a unique ray trajectory passing across the point of
coordinate ${\mathbf{x}}$, we say that $\Phi$ has a critical nondegenerate
point at $s=s_0$. The surfaces $\Phi={\mbox{Const.}}$ represent the constant phase
surfaces. Next, recalling the Morse Lemma, which refers to the representation of
functions all of whose critical points are nondegenerate, we can obtain an
asymptotic evaluation, for large $k$, of integral (\ref{2.6}) in the following
form \cite{Fedoryuk}:
\begin{equation}
\psi({\mathbf{x}};k)\,\simeq\,e^{{\rm i} k\Phi({\mathbf{x}};s_0)}
\sum_{m=0}^\infty\frac{A_m({\mathbf{x}})}{({\rm i} k)^m}.
\label{2.7}
\end{equation}
The leading term of expansion (\ref{2.7}) reads
\begin{equation}
\psi({\mathbf{x}};k) \simeq A_0({\mathbf{x}})\, e^{{\rm i} k
\Phi({\mathbf{x}};s_0)},
\label{2.8}
\end{equation}
where
\begin{equation}
A_0({\mathbf{x}})=A({\mathbf{x}};s_0)\left(\left|
\frac{\partial^2\Phi}{\partial s^2}\right|^{-1/2}\right)_{s=s_0}
\exp\left[{\rm i}\,\frac{\pi}{4}{\,\,{\mbox{sgn\,}}}\left(
\frac{\partial^2\Phi}{\partial s^2}\right)_{s=s_0}\right].
\label{2.9}
\end{equation}
By substituting the leading term (which, for simplicity, will be
written as $A\exp({\rm i} k\Phi)$) into equation (\ref{2.1}), collecting the
powers of $({\rm i} k)$, and, finally, equating to zero their coefficients,
two equations are obtained: the eikonal (or Hamilton--Jacobi) equation
\begin{equation}
g^{ij}\frac{\partial\Phi}{\partial x_i}\, \frac{\partial\Phi}{\partial x_j}
\equiv g^{ij} p_i p_j = 1,
\label{2.10}
\end{equation}
\noindent
and the transport equation
\begin{equation}
\frac{1}{\sqrt g}\, \sum_{i=1}^n\frac{\partial}{\partial x_i} \left[\sqrt{g}
\left(A^2\sum_{j=1}^n g^{ij}\,\frac{\partial\Phi}{\partial x_j}\right)\right]=0,
\label{2.11}
\end{equation}
\noindent
whose physical meaning is the conservation of the current density.
Recalling that in the case of orbiting bound states the manifold
${\mathbb{M}}^n$ is the unit sphere ${\mathbb{S}}^2$, and assuming that the
phase $\Phi$ and the amplitude $A$ do not depend on the angle $\varphi$,
from (\ref{2.10}) and (\ref{2.11}) we obtain
\begin{eqnarray}
&&\left(\frac{d\Phi}{d\theta}\right)^2=1, \label{2.12} \\
&&\frac{1}{|\sin\theta|}\left[\frac{d}{d\theta}\left(|\sin\theta|\: A^2 \:
\frac{d\Phi}{d\theta}\right)\right]=0 \qquad (\theta \neq m\pi, m\in{\mathbb{Z}}).  \label{2.13}
\end{eqnarray}
It follows from (\ref{2.12}) that $\Phi = \pm\theta+{\mbox{Const.}}$, whereas
(\ref{2.13}) implies $A(\theta)={\mbox{Const.}}\cdot|\sin\theta|^{-1/2}$,
$\theta\neq n\pi$ ($n\in{\mathbb{Z}}$). Therefore, the approximation (\ref{2.8}) becomes
\begin{equation}
\psi(\theta;k)=\frac{C}{\sqrt{|\sin\theta|}}\,
e^{\pm{\rm i} k\theta}\qquad(\theta\neq n\pi, n\in{\mathbb{Z}},\, C = {\rm constant}),
\label{2.14}
\end{equation}
where the terms $\exp(\pm{\rm i} k\theta)$ represent waves travelling in
counterclockwise ($\exp({\rm i} k\theta)$) or clockwise
($\exp(-{\rm i} k\theta)$) sense around the unit sphere.
Since approximation (\ref{2.14}) fails at
$\theta=n\pi$ ($n\in{\mathbb{Z}}$), we must consider
the problem of patching these approximations when the surface rays cross the
antipodal points $\theta=0,\pi$; these are conjugate points, and the Morse
Index Theorem can be applied. Accordingly, the uniformization of approximation
(\ref{2.14}) consists precisely in finding the phase--shift
associated with the crossing of rays through the antipodal conjugate points
$\theta=0$ and $\theta=\pi$.

\subsection{Uniformization of the eikonal approximation and geometric quantization}
\label{subse:uniformization}
It follows from Maslov theory \cite{Maslov} that in any domain in which the
configuration space representation of a Lagrangian manifold fails, one has the
possibility of working out the problem in a momentum space representation.
While the intensity of the ray tube, which is proportional to $(\sin\theta)^{-1}$,
is infinite in the
neighborhood of the conjugate points, the intensity is instead finite in the
$p_\theta$--representation \cite{Maslov}. Therefore, a semiclassical approximation
to the wavefunction in almost any singular region can be obtained by means of a
transformation to a suitably chosen momentum space, or mixed space
representation. Once the local asymptotic solution in terms of the
mixed $(p_\theta,\varphi)$ representation has been computed, it is possible
to return to the configuration space $(\theta,\varphi)$ by transforming
$p_\theta\rightarrow\theta$ through the inverse Fourier transformation
$F^{-1}_{k,p_\theta\rightarrow\theta}$. In this way one determines the
so--called Maslov indeces, which allow us to obtain the uniformization
of approximation (\ref{2.14}) across the conjugate points.

Here we shall follow another approach based on the analytic continuation. With
this in mind, we prove the following proposition.
\begin{proposition}
\label{pro:2}
{\rm (i)} The geometrical approximation (\ref{2.14}) (eikonal approximation) can be
extended to include the crossing of the conjugate points by adding a
phase--shift $(-\frac{\pi}{2})$ ($\frac{\pi}{2}$, respectively) at each
counterclockwise (clockwise) crossing of each antipodal conjugate point,i.e.,
\begin{itemize}
\item[\rm (a)] at each counterclockwise crossing, one has the shift
\begin{equation}
e^{{\rm i} k\theta} \longrightarrow e^{{\rm i} (k\theta-\pi/2)};
\label{2.15}
\end{equation}
\item[\rm (b)] at each clockwise crossing, one has the shift
\begin{equation}
e^{-{\rm i} k\theta} \longrightarrow e^{-{\rm i} (k\theta-\pi/2)}.
\label{2.16}
\end{equation}
\end{itemize}
{\rm (ii)} The orbital angular momentum $L$ can assume only half--integer values
according to the rule
\begin{equation}
L = \ell + \frac{1}{2}~~~~~(\ell=0,1,2,\ldots; \hbar = 1).
\label{2.17}
\end{equation}
{\rm (iii)} The asymptotic representation of the angular orbiting bound state
wavefunction $\psi_\ell(\theta,k)$ where $\ell=0,1,2,\ldots$, is given by
\begin{equation}
\psi_\ell(\theta,k)\simeq\frac{C}{\sqrt{|\sin\theta|}}\sin
\left[\left(\ell+\frac{1}{2}\right)\theta+\frac{\pi}{4}\right]
\qquad (\theta\neq n\pi, n\in{\mathbb{Z}}).
\end{equation}
\end{proposition}

\noindent
{\bf Proof:} (i) Let us consider the function $(\sin\theta)^{-1/2}$, i.e., the
amplitude of the eikonal approximation (\ref{2.14}) up to a constant. This
function presents branch singularities at $\theta=n\pi$ ($n\in{\mathbb{Z}}$).
The continuation can be simply realized by overpassing these singularities by
semicircular arcs, which are oriented counterclockwise for
$\theta\in [0,+\infty)$, and oriented clockwise for $\theta\in(0,-\infty)$.
Now, rewrite approximation (\ref{2.14}) in the following form:
$\psi(\theta,k)=C \exp(\log\!A \pm {\rm i} k\theta)$ ($A=(\sin\theta)^{-1/2}$,
$C={\mbox{Const.}}$). When the wavefunction $\psi$ is varied across a branch
point in the counterclockwise direction, the term $\log\!A$ acquires a factor
$-{\rm i}\frac{\pi}{2}$; analogously, when $\psi$ is varied across a branch
point moving in clockwise direction, the term $\log\!A$ acquires a factor
${\rm i}\frac{\pi}{2}$. Thus, the uniformization of the wavefunction
approximation (\ref{2.14}) can be achieved by patching the rays travelling
along the circle ${\mathbb{S}}^1$ as follows:
add a phase--shift of $-\frac{\pi}{2}$ to the phase of the wave travelling
in counterclockwise direction as the ray passes through each antipodal
conjugate point whose relative distance on the unit sphere is $\pi$.
Thus, the rule (\ref{2.15}) is obtained. Analogously, add a phase--shift of
$\frac{\pi}{2}$ to the phase of the wave travelling in clockwise direction as
the ray passes through each antipodal conjugate point, thus obtaining
the rule (\ref{2.16}). \\
(ii) We now evaluate the variation of the phase of the wavefunction
$\psi(\theta,k)$ associated with a circular orbit $\gamma^{(+)}$ ($0\leqslant\theta\leqslant 2\pi$)
oriented counterclockwise. In order to
guarantee the single--valuedness of the wavefunction $\psi$, we impose the
following condition (see also \cite{Keller2}):
\begin{equation}
\Delta_{\gamma^{(+)}(0,2\pi)}\left(\log\!A\right)+{\rm i}
\Delta_{\gamma^{(+)}(0,2\pi)}(k\theta)={\rm i} 2\pi\ell \qquad (\ell=0,1,2,\ldots).
\label{2.18}
\end{equation}
Since $\Delta_{\gamma^{(+)}(0,2\pi)}(\log\!A)=-{\rm i}\pi$, and $\theta=2\pi$ when the
circular orbit $\gamma^{(+)}$ is completed, it follows from formula (\ref{2.18}) that
\begin{equation}
-\frac{1}{2}+\Delta_{\gamma^{(+)}(0,2\pi)}(k)=\ell \qquad (\ell=0,1,2,\ldots).
\label{2.19}
\end{equation}
Recalling that we are considering, for simplicity, the orbiting around a unit
sphere, it follows, from formula (\ref{2.19}) that the values of the angular
momentum compatible with the single--valuedness of the wavefunction
$\psi$ are given by: $L=\ell+\frac{1}{2}$ $(\ell=0,1,2,\ldots;\hbar=1)$,
i.e., by formula (\ref{2.17}).

\vspace{2.5ex}

\noindent
{\bf Remark:} Let us note that the universal covering
$p(t):{\mathbb{R}}^1\rightarrow{\mathbb{S}}^1$, $p(t)=e^{2\pi{\rm i} t}$, is
generated by a properly discontinuous transformation group
by the translations $t\rightarrow t+n$ ($n\in{\mathbb{Z}}$) of the axis
${\mathbb{R}}^1$, namely, the exponential map wraps the line ${\mathbb{R}}^1$ around
the unit circle in ${\mathbb{R}}^2$. We can thus relate the semiclassical
quantization of the orbital angular momentum $L$ (formula (\ref{2.17})) to the
topological properties of the fundamental group
$\pi_1({\mathbb{S}}^1) \simeq {\mathbb{Z}}$.

\vspace{2.5ex}

\noindent
(iii) Let $\psi_\ell^{(+)}$ and $\psi_\ell^{(-)}$ denote the contribution to the
angular wavefunction $\psi_\ell$ given by the counterclockwise and clockwise
travelling waves, respectively. The total wavefunction $\psi_\ell$ is the sum
$\psi_\ell^{(+)}+\psi_\ell^{(-)}$. Taking the statements (i) and (ii) into account,
at a chosen $\ell$, we have (see formula (\ref{2.17})):
\begin{eqnarray}
\lefteqn{\psi_\ell(\theta,k)=\psi_\ell^{(+)}(\theta,k)+\psi_\ell^{(-)}(\theta,k)=
C(k)\,A(\theta)\left[e^{{\rm i} k\theta}+{\rm i} e^{{\rm i} k(2\pi-\theta)}\right]}
\nonumber \\
& & \mbox{}=2C(k)A(\theta)\,e^{{\rm i} k\pi}e^{{\rm i}\pi/4}
\left[\frac{e^{{\rm i}[k(\theta-\pi)-\pi/4]}+
e^{-{\rm i}[k(\theta-\pi)-\pi/4]}}{2}\right] \nonumber \\
& & \mbox{}=2\,(-1)^\ell\,e^{{\rm i}3\pi/4}C(k)A(\theta)\cos\left[k(\theta-\pi)-\pi/4\right].
\label{2.20}
\end{eqnarray}
Since $k=\ell+\frac{1}{2}$, we obtain
\begin{eqnarray}
\lefteqn{\psi_\ell(\theta,k)=\psi_\ell^{(+)}(\theta,k)+\psi_\ell^{(-)}(\theta,k)}
\nonumber \\
& & \mbox{} = \frac{C_1(k)}{\sqrt{|\sin(\theta-\pi)|}}\cos
\left[\left(\ell+\frac{1}{2}\right)(\theta-\pi)-
\frac{\pi}{4}\right] \qquad (\theta\neq n\pi, n\in{\mathbb{Z}}),
\label{2.20bis}
\end{eqnarray}
which is proportional to the well--known asymptotic behavior of the Legendre polynomials
for large values of $\ell$.
\hfill$\Box$

Typical examples of phenomena which can be theoretically described in terms of
semiclassical orbiting are the rotational resonances present in molecular
scattering. These quasi--bound states have been observed with particularly
clear evidence in the H--Kr collision. In \cite{Toennies}, a phase--shift
analysis of these experimental scattering data is presented. The energy values
corresponding to the upward crossing through $\frac{\pi}{2}$ of these
phase--shifts give the energy locations of the resonance peaks.
In \cite{DeMicheli2}, these values of the center of mass energy
are reported in correspondence with the integer values of $\ell$ ($\ell=4,5,6$).
Then, these points are interpolated with the line $\ell(\ell+1)$ as a function
of the center of mass energy $E$. The good agreement between the energy
locations $(E>0)$ of the resonance peaks and the line $\ell(\ell+1)$ strongly
supports the correctness of the linear parametrization
\begin{equation}
\ell(\ell+1) = 2IE + c_0,
\label{n1}
\end{equation}
where $I=\mu R^2$ is the moment of inertia of the two--particle system, $\mu$ is
the reduced mass, $R$ is the interparticle distance, $E$ is the center of mass
energy, and $c_0$ is a constant. Formula (\ref{n1}) proves that these resonances
are generated by the rotational dynamics. Since the intercept $c_0$ turns out to
be larger than zero (see Fig. 2 of \cite{DeMicheli2}), the straight
trajectory interpolating the resonances for $E>0$ can be extrapolated
to $E<0$, and then the energy values of the bound states corresponding to
$\ell=0,1,2,3$ can be determined
(see once again Fig. 2 of \cite{DeMicheli2}).
Formula (\ref{2.17}) gives the semiclassical approximation of the
quantum--mechanical angular momentum $\sqrt{\ell(\ell+1)}$ ($\hbar=1$).
Let us finally note that the theoretical results obtained in this section refer
strictly to the bound states phenomena ($E<0$). The resonances $(E>0)$ will be
reconsidered in Subsection \ref{subse:orbiting}, taking into account, in particular,
the lifetime of these states, which is closely related to the tunnelling across
the centrifugal barrier.

\section{Riemannian obstacle problem and the Ludwig system}
\label{se:riemannian}
\subsection{Existence of diffracted rays and nonuniqueness of the Cauchy
problem for a Riemannian manifold with boundary}
\label{subse:non}
Let us now consider the so--called {\it Riemannian obstacle problem}
\cite{Alexander1}. A boundary component is viewed as an obstacle around which a
geodesic can bend, or at which a geodesic can end. Let ${\mathbb{K}}$ denote a
convex obstacle which is embedded in a complete $n$--dimensional manifold
${\mathbb{M}}^n$ ($n\geqslant 3$) equipped with a Riemannian metric,
where $g_{ij}$ stands for the metric tensor. Let us introduce the space
${\mathbb{M}}^n\backslash\overline{{\mathbb{K}}}\equiv{\mathbb{M}}^n_{\mathbb{K}}$,
where the obstacle ${\mathbb{K}}$ is an open connected subset of ${\mathbb{M}}^n$,
with regular boundary $\partial{\mathbb{K}}$ and compact closure
$\overline{{\mathbb{K}}}={\mathbb{K}} \cup \partial{\mathbb{K}}$. In the
following, we shall denote the space ${\mathbb{M}}^n_{\mathbb{K}}$ simply by
${\mathbb{M}}$. We are thus led to consider the space
${\mathbb{M}}^*={\mathbb{M}} \cup \partial{\mathbb{K}}$, which has the structure
of a manifold with boundary. In order to remedy the lack of geodesic
completeness, i.e., the possibility of extending every geodesic infinitely
and in a unique way, one can introduce the notion of {\it geodesic terminal}
\cite{Plaut} to represent a point where a geodesic stops. Indeed, following
Plaut \cite{Plaut}, it can be proved that ${\mathbb{M}}^*$ is the completion
of ${\mathbb{M}}$ by observing that the set ${\mathcal{I}}$ of the geodesic
terminals is nowhere dense, and
$\partial{\mathbb{M}}^*=\partial{\mathbb{K}}={\mathcal{I}}$.

It is convenient to model the manifold with boundary on the half--space
${\mathbb{R}}^n_+=\{(x_1,\ldots,x_n)\in{\mathbb{R}}^n | x_n \geqslant 0\}
\subset {\mathbb{R}}^n$, where ${\mathbb{R}}^{n-1}_0$ stands for the boundary
$x_n=0$ of ${\mathbb{R}}^n_+$. Thus, for a manifold with boundary,
there exists an atlas $\{{\mathcal{U}}^\alpha\}$, with local coordinates
$(x_1^\alpha,\ldots,x_n^\alpha)$, such that, in any chart, we have the strict
inequality $x_n^\alpha > 0$ at the interior points, and $x_n^\alpha = 0$ at the
boundary points \cite{Mishchenko}. The set $\partial{\mathbb{M}}^*$ of the
boundary points is a smooth manifold of dimension $(n-1)$. In particular,
it is convenient to introduce suitable coordinates $x_i^{\rm b}$ $(i=1,\ldots,n)$
adapted to the boundary, the so--called {\it geodesic boundary coordinates}
\cite{Alexander1}, with $x_n^{\rm b}$ defined as the distance from the
boundary $\partial{\mathbb{M}}^*$; note that $x_n^{\rm b}$ satisfies necessarily
a positivity condition, i.e., $x_n^{\rm b} \geqslant 0$. Then, starting with
arbitrary coordinates $x_i^{\rm b}$, $i<n$ on $\partial{\mathbb{M}}^*$, one
extends them to be constant  on ordinary geodesics normal to
$\partial{\mathbb{M}}^*$. These coordinates have been used indeed by
Alexander, Berg and Bishop (ABB) in order to write the equation of a geodesic
$\gamma$ of ${\mathbb{M}}^*$ \cite{Alexander1}. Let us in fact note that the geodesics in a
Riemannian manifold without boundary are the solutions of a system
of differential equations with suitable Cauchy conditions, and the main result
is the theorem which guarantees existence, uniqueness and smoothness of the
solution of the Cauchy problem of this system. Conversely, the case of a
Riemannian manifold with boundary is different from the classical one
concerning uniqueness and smoothness of the solution. In \cite{DeMicheli1}, the
following proposition, which is essentially due to (ABB), has been proved.
\begin{proposition}
\label{pro:3}
{\rm (i)} In a Riemannian manifold with boundary (the normal curvature of the boundary
being different from zero) the determination of the geodesics by their initial
tangents (Cauchy problem) is not unique.\\
{\rm (ii}) In any point of the boundary $\partial{\mathbb{K}}$
($\partial{\mathbb{K}}$ convex) we have
\begin{itemize}
\item[\rm (a)] a geodesics of class $C^\infty$ bending around the boundary;
\item[\rm (b)] a geodesics which leaves tangentially the boundary
and which is of class $C^1$ at the point of contact with the boundary.
\end{itemize}
\end{proposition}
The diffracted rays are precisely the geodesics of ${\mathbb{M}}^*$ obtained by
gluing together a geodesic segment belonging to the boundary (whose normal
curvature is supposed to be different from zero) with a geodesic segment
belonging to the interior of ${\mathbb{M}}^*$. They are of class $C^1$ at the
transition point. If we assume that the space ${\mathbb{M}}^n$ in which the
obstacle is embedded is ${\mathbb{R}}^3$ (equipped with an Euclidean metric)
then, before and beyond the obstacle, the diffracted rays consist of straight lines
tangent to the obstacle. Therefore, as a consequence of the
nonuniqueness of the Cauchy problem at the boundary of the manifold, we can
conclude by stating that at each point of the boundary we have a bifurcation;
the ray splits in two parts, where one part continues as an ordinary straight line ray
(diffracted ray of class $C^1$ at the splitting point) and the other part travels
along the surface as a geodesic of the boundary.

\subsection{Morse Index Theorem for Riemannian manifolds with boundary and
homotopic classes of diffracted rays}
\label{subse:morse}
From now on, as a typical example of obstacle, we regard the spherical ball of unit
radius embedded in ${\mathbb{R}}^3$: $\partial{\mathbb{K}}={\mathbb{S}}^2$,
equipped with the standard metric. However, in view of the homotopic invariance,
the results which we derive hold for any convex and compact obstacle
represented by Besse--type manifolds (all of whose geodesics are closed) having
a symmetry axis. Let us denote by ${{\mathcal{A}}}$ the {\it extended--axis} of
symmetry of the obstacle, passing through the point
$p_0 \in {\stackrel{\scriptscriptstyle\hspace{-.9ex}\circ}{\mathbb{M}^*}}$, at which
the source of light is located. Let ${{\mathcal{A}}}_-$ denote the portion of
this axis lying in the illuminated region (i.e., the same side of the light
source), and ${{\mathcal{A}}}_+$ the portion of the axis lying in the shadow.
The axis ${{\mathcal{A}}}$ equals
${{\mathcal{A}}}_- \cup D \cup {{\mathcal{A}}}_+$, where $D$ is the diameter
of the sphere. Let us note that all the points of ${\mathbb{M}}^*$ that do not
belong to ${{\mathcal{A}}}_+$ are connected to $p_0$ by only one geodesic of
minimal length, whereas the points $q \in {{\mathcal{A}}}_+$ are connected to
$p_0$ by a continuum of geodesics of minimal length that can be obtained as the
intersections of ${\mathbb{M}}^*$ with planes passing through the axis
${{\mathcal{A}}}$. By rotating these planes, and keeping $p_0$ and $q$ fixed,
we obtain a variation vector field which is a Jacobi field vanishing at $p_0$
and $q$. Thus, we can conclude that $p_0$ and $q$ are conjugate with multiplicity
one, because the possible rotations are along one direction only. We are then
led to pose the question weather the Morse Index Theorem can be extended to
include Riemannian manifolds with boundary. To this end, we introduce the
so--called {\it regular geodesics} in the sense of Alexander \cite{Alexander2}.
Following \cite{Alexander2}, a geodesic $\gamma$ is regular if
\begin{itemize}
\item[(a)] all boundary contact intervals of $\gamma(t)$ have positive measure;
\item[(b)] the points of arrival of $\gamma(t)$ at $\partial{\mathbb{M}}^*$ are
not conjugate to the initial point $\gamma(0)$.
\end{itemize}
It is easy to show that the diffracted rays are regular geodesics. It is then
possible to define a Hessian form which is simply given by the sum of the
classical formulae for $\partial{\mathbb{M}}^*$ on contact intervals and of the
formulae for the interior of ${\mathbb{M}}^*$ (i.e.,
${\stackrel{\scriptscriptstyle\hspace{-.9ex}\circ}{\mathbb{M}^*}}$) on interior
intervals. We can thus formulate the extended Morse Index Theorem as follows.
\begin{proposition}{{\rm(Morse Index Theorem for Riemannian Manifolds with
Boundary \cite{Alexander2})}}
\label{pro:4}
Let $\gamma$ be a regular geodesic; then the index $\mu$ of $S_{**}$ is finite
and equal to the number of points in $\gamma(t)$ conjugate to the
initial point $\gamma(0)$ $(0\leqslant t \leqslant 1)$ and counted according to their
multiplicities.
\end{proposition}

\begin{figure}[ht]
\centering
\includegraphics[width=4in]{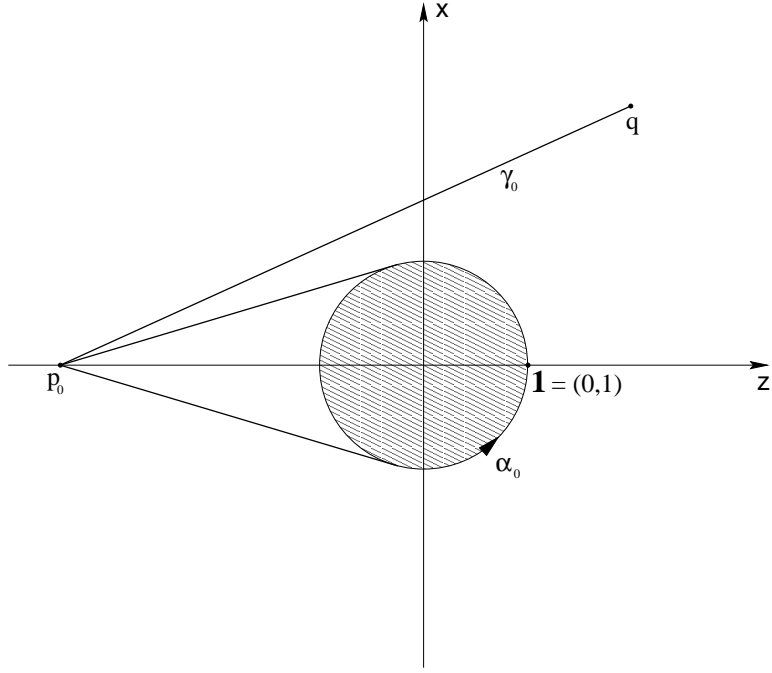}
\caption{\label{fig_1}\small Geometry of the problem in the case of spherical obstacle.}
\end{figure}

Next we focus on the points
$q \in {\stackrel{\scriptscriptstyle\hspace{-.9ex}\circ}{\mathbb{M}^*}}
\backslash ({{\mathcal{A}}}_+ \cup {{\mathcal{A}}}_-)$
(i.e., the points belonging to the interior of ${\mathbb{M}}^*$ that do not lie on the
axis of the obstacle). Consider the space $X_{p_0 q}=(E^2_q)\cap{\mathbb{M}}^*$,
where $(E^2_q)$ is the Euclidean plane uniquely determined by the axis of the
obstacle through $p_0$ and the point $q$. The space $X_{p_0 q}$ (which will be denoted
hereafter simply by $X$) is arcwise connected, and its boundary in
$(E^2_q)$ is a circle ${\mathbb{S}}^1$, which is a deformation retract of $X$.
In view of these facts, the fundamental group $\pi_1(X;p_0)$ does not depend on
the base point $p_0$ and is isomorphic to $\pi_1({\mathbb{S}}^1;{\bf 1})$,
where ${\bf 1}$, which is described in a convenient system by the coordinates $(0,1)$
(see Fig. \ref{fig_1}), represents the contact point of
${{\mathcal{A}}}_+$ with ${\mathbb{S}}^2$,
$\pi_1(X;p_0) \sim \pi_1({\mathbb{S}}^1;{\bf 1}) \simeq {\mathbb{Z}}$.
Let us consider, within the fundamental grupoid $\Pi_1(X)$ of $X$, the set
$\Pi_1(X;p_0,q)$ of the paths in $X$ connecting $p_0$
with $q$, modulo homotopy with fixed endpoints. Let us clarify our choices. \\
(i) We introduce in $(E^2_q)$ ($q$ fixed in
$X\backslash ({{\mathcal{A}}}_+ \cup {{\mathcal{A}}}_-)$)
an orthogonal reference system, whose origin coincides with the centre of the
spherical ball representing the obstacle, and the {\it extended axis} of
symmetry ${{\mathcal{A}}}$ coincides with the horizontal axis. Recall that the
source is located in the point $p_0\in {{\mathcal{A}}}_-$, while
the point $q$, exterior to ${{\mathcal{A}}}$, lies in the upper half--plane of
this reference system, i.e., in the first or in the second quadrant; \\
(ii) $\gamma_0$ is the geodesic from $p_0$ to $q$ of minimal length; \\
(iii) $\alpha_0$ is a loop in $X$ at $p_0$ such that: \\
\indent $[\alpha_0]$ is a generator of $\pi_1(X;p_0)$ (establishing an isomorphism
$\pi_1(X;p_0) \simeq {\mathbb{Z}}$); \\
\indent $\alpha_0$ turns in counterclockwise sense around the obstacle
(see Fig. \ref{fig_1}).

\noindent
We can now state the following proposition.
\begin{proposition}
\label{pro:5}
{\rm (i)} Each element of $\pi_1(X;p_0)$ is a homotopy class $[\alpha]$, with fixed
endpoints, of a certain loop $\alpha$ in the space $X$, starting and ending at
the point $p_0$. \\
{\rm (ii)} Each path $c_0$ from $p_0$ to
$q \in {\stackrel{\scriptscriptstyle\hspace{-.9ex}\circ}{\mathbb{M}^*}}
\backslash ({{\mathcal{A}}}_+ \cup {{\mathcal{A}}}_-)$
determines a one--to--one correspondence $W$ between $\pi_1(X;p_0)$ and the set
$\Pi_1(X;p_0,q)$. Such a correspondence can be constructed as:
$\forall [c] \in \Pi_1(X;p_0,q):[c]\rightarrow[c\star c_0^{-1}]\in\pi_1(X;p_0)$,
where the symbol '$\star$' denotes the concatenation of paths, and $c_0^{-1}$
denotes the reverse path of $c_0$. \\
{\rm (iii)} In each homotopy class of $\pi_1(X;p_0)$, there is precisely one element
of minimal length. In each homotopy class of
$\Pi_1(X;p_0,q)$ ($q \in
{\stackrel{\scriptscriptstyle\hspace{-.9ex}\circ}{\mathbb{M}^*}}
\backslash ({{\mathcal{A}}}_+ \cup {{\mathcal{A}}}_-)$)
there is precisely one geodesic $\gamma$. \\
{\rm (iv)} The correspondence $W$ assigns to each geodesics $\gamma$ the number
$w(\gamma)$ defined by $[\gamma \star \gamma_0^{-1}]=w(\gamma)[\alpha_0]$;
$w(\gamma)$ is the winding number of the loop $[\gamma \star \gamma_0^{-1}]$
determined by $\gamma$ (with respect to the chosen generator $[\alpha_0]$). \\
{\rm (v)} A bijective correspondence can be established between the winding
number $w(\gamma)$ of the loop $[\gamma \star \gamma_0^{-1}]$ ($\gamma_0$
being the geodesic of minimal length connecting $p_0$ and $q$), and the
crossing number $n(\gamma)$ which counts the intersections of the geodesics
$\gamma$ with the horizontal axis (i.e., the extended symmetry axis). The
winding number determines the crossing number through the following bijective
correspondence:
\begin{center}
\begin{tabular}{cccc}
${\mathbb{Z}}$ & $\longrightarrow$ & ${\mathbb{N}}$ & \\
$n$  & $\longrightarrow$ & $(2n+1)$ & if ~~$n \geqslant 0,$ \\
$n$  & $\longrightarrow$ & $-2n$ & if ~~$n < 0.$
\end{tabular}
\end{center}
\end{proposition}
It follows from this proposition that the subset
${{\mathcal{A}}}_+ \cup {{\mathcal{A}}}_-$ of the {\it extended symmetry axis}
of the obstacle ${\mathbb{K}}$ (including the antipodal points of the
spherical ball belonging respectively to ${{\mathcal{A}}}_-$ and
${{\mathcal{A}}}_+$) forms the set of conjugate points, which we can properly
refer to as the {\it axial caustic}. Accordingly, in view of the extended Morse Index
Theorem, the index $\mu$ of the Hessian jumps by one when the geodesic
(i.e., the ray trajectory) crosses the caustic.

\begin{figure}[tb]
\centering
\includegraphics[width=4in]{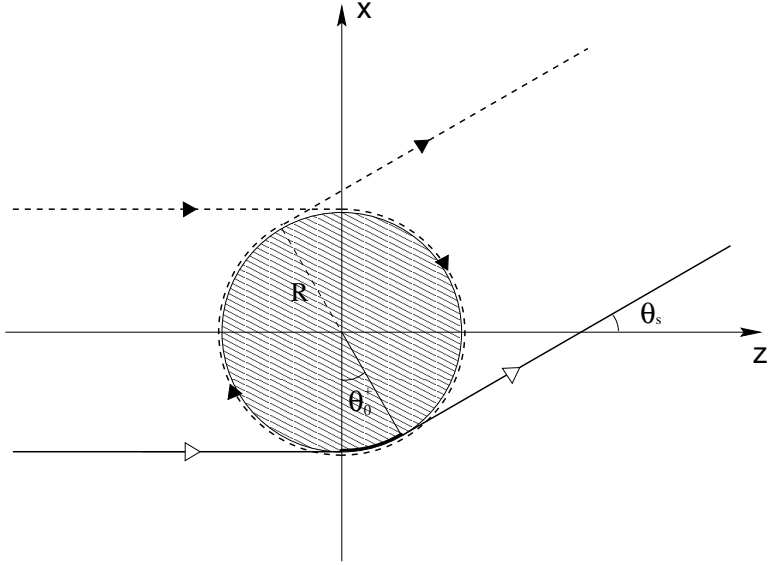}
\caption{\label{fig_2}\small Diffractive scattering: geometry of the
contribution of two grazing rays to the scattered amplitude.}
\end{figure}

\subsection{Geometric structure of the caustics}
\label{subse:geometric}
We can prove the following proposition.
\begin{proposition}
\label{pro:new6}
If the obstacle is a spherical ball of radius $R$ and the extended symmetry axis of the ball
is given by ${\mathcal{A}}={\mathcal{A}}_-\cup D \cup {\mathcal{A}}_+$, then the
caustics are as follows:
\begin{itemize}
\item[\rm (a)] the surface of the spherical ball, which is the envelope of the diffracted rays;
\item[\rm (b)] the semiaxes ${\mathcal{A}}_+$ and ${\mathcal{A}}_-$, which form the axial caustic.
\end{itemize}
\end{proposition}

\noindent
{\bf Proof:}
Let us introduce an orthogonal system of axes $(X,Y,Z)$ in ${\mathbb{R}}^3$
whose origin coincides with the center of the obstacle and such that the
$Z$--axis (whose corresponding coordinate is denoted by $z$), chosen parallel
to the incoming beam of rays, is positively oriented in the direction of the
outgoing rays (see Fig. \ref{fig_2}). Indeed, we assume that the point $p_0$,
where the source is located, is pushed to $-\infty$. Then we introduce, in
addition, the following coordinates on the sphere: ${\mathbf{r}}_0$ is the radial vector,
$\varphi_0$ is the azimuthal angle, $\theta_0$ is the angle measured along the
meridian circle from the point of incidence of the ray trajectory on the sphere.
We start by considering the rays that leave the surface of the sphere after
diffraction. Let us recall that the interior of the space ${\mathbb{M}}^*$
(i.e., ${\stackrel{\scriptscriptstyle\hspace{-.9ex}\circ}{\mathbb{M}^*}}$),
exterior to the obstacle, can be regarded as an Euclidean space without boundary.
Next, we switch from the wave to the ray description, writing the eikonal equation
$g^{ij}\frac{\partial \Phi}{\partial x_i}\frac{\partial \Phi}{\partial x_j}=1$ in Hamiltonian form,
where $\Phi$ stands for the phase function, $x_i$ $(i=1,2,3)$ are the Cartesian
coordinates which refer to the axes $(X,Y,Z)$: $x_1=x,x_2=y,x_3=z$.
We then set $p_i=\frac{\partial \Phi}{\partial x_i}$ and introduce the
function $F=\frac{1}{2}(g^{ij}p_ip_j-1)$. The Hamilton characteristic system
reads:
\begin{subequations}
\label{z2}
\begin{eqnarray}
\frac{d x_i}{d\tau} &=& g^{ij}p_j, \label{z2a} \\
\frac{d p_i}{d\tau} &=& -\frac{\partial F}{\partial x_i}=
-\frac{1}{2}p_ip_j\frac{\partial g^{ij}}{\partial x_i},
\label{z2b}
\end{eqnarray}
\end{subequations}
$(i,j=1,2,3)$, where $\tau$ is a running parameter along the ray emerging from
the surface $\Phi={\mbox{Const.}}$. By (\ref{z2a}), we have
\begin{equation}
\label{z3}
d\Phi=\sum_{i=1}^3\frac{\partial \Phi}{\partial x_i}\,dx_i=
\sum_{i=1}^3 p_i\,dx_i =
g^{ij}p_ip_j\,d\tau=d\tau,
\end{equation}
where the last equality follows from $g^{ij}p_ip_j=1$. By Eqs. (\ref{z2}) and
recalling that ${\stackrel{\scriptscriptstyle\hspace{-.9ex}\circ}{\mathbb{M}^*}}$
is Euclidean, we obtain
\begin{subequations}
\label{z5}
\begin{eqnarray}
\frac{d {\mathbf{r}}}{d\tau} &=& {\mathbf{p}}, \label{z5a} \\
\frac{d {\mathbf{p}}}{d\tau} &=& 0, \label{z5b}
\end{eqnarray}
\end{subequations}
where ${\mathbf{r}}=(x_1,x_2,x_3)$, and ${\mathbf{p}}=\nabla\Phi$. Then it follows from (\ref{z5}) that
\begin{equation}
\label{z6}
{\mathbf{r}}(\theta_0,\varphi_0,\tau)={\mathbf{r}}_0(\theta_0,\varphi_0)+
\tau\,{\mathbf{p}}_0(\theta_0,\varphi_0),
\end{equation}
where ${\mathbf{p}}_0$ is the unit vector tangent to the obstacle where the ray leaves
the sphere.

Consider now the ray hitting the sphere at the point of coordinates $(-R,0,0)$
and then travelling in the counterclockwise direction. The components of the vectors
${\mathbf{r}}_0$ and ${\mathbf{p}}_0$ are (see Fig. \ref{fig_2}):
\beq
{\mathbf{r}}_0=(-R\cos\theta_0\cos\varphi_0,R\cos\theta_0\sin\varphi_0,R\sin\theta_0),
\label{coord1}
\eeq
and
\beq
{\mathbf{p}}_0=(\sin\theta_0\cos\varphi_0,-\sin\theta_0\sin\varphi_0,\cos\theta_0).
\label{coord2}
\eeq
Substituting these expressions in (\ref{z6}) gives
\begin{eqnarray}
{\mathbf{r}}(\theta_0,\varphi_0,\tau)&=&
(-R\cos\theta_0\cos\varphi_0+\tau\sin\theta_0\cos\varphi_0, \nonumber \\
& & R\cos\theta_0\sin\varphi_0-\tau\sin\theta_0\sin\varphi_0,
R\sin\theta_0+\tau\cos\theta_0).
\label{2.23}
\end{eqnarray}
Then the Jacobian determinant $\left|\frac{\partial(x,y,z)}{\partial(\theta_0,\varphi_0,\tau)}\right|$
of the transformation $(x,y,z)\leftrightarrow (\theta_0,\varphi_0,\tau)$ is
\begin{equation}
|J| = \tau(R\cos\theta_0-\tau\sin\theta_0).
\label{2.24}
\end{equation}
It vanishes for $\tau=0$ and for $\tau=\overline{\tau}\equiv R\cot\theta_0$.
Analogous calculations can be performed for the rays oriented clockwise.
Finally, in order to distinguish the two orientations of the rays, we add the
superscript $(+)$ to everything referring to the counterclockwise oriented rays, and
the superscript $(-)$ when referring to the clockwise oriented rays.
In conclusion, we see that the domain on which the Jacobian determinant vanishes is formed by
\begin{itemize}
\item[(i)] the surface of the spherical ball corresponding to $\tau^\pm=0$;
\item[(ii)] the semiaxes represented by
\begin{equation}
\tau^\pm = \overline{\tau}^\pm \equiv R\cot\theta_0^\pm:\left\{\begin{array}{l}
\mbox{\rm the semiaxis}~{{\mathcal{A}}}_+ ~{\rm if} ~0\leqslant\theta_0^\pm\leqslant\pi, \\[+5pt]
\mbox{\rm the semiaxis}~{{\mathcal{A}}}_- ~{\rm if} ~\pi\leqslant\theta_0^\pm\leqslant 2\pi.
\end{array}\right.
\label{new29}
\end{equation}
\end{itemize}
\hfill$\Box$

\subsection{Chester, Friedmann and Ursell approximation and the Ludwig system}
\label{subse:chester}
Let us now consider the caustic generated by the boundary.
As we have seen in Proposition \ref{pro:3}, at any point of the boundary, the ray
splits into two parts: one part leaves the boundary along the tangent; the other one
travels along the surface and bends around the obstacle. Therefore, in the domain at the
exterior of the boundary but close to it, the Lagrangian manifold is composed of
all the trajectories tangent to the boundary $\partial{\mathbb{K}}$
that have $\partial{\mathbb{K}}$ as an envelope. Each point $P$ in this domain lies
in the intersection of two diffracted rays which are tangent to the border of the
diffracting body and cross two surfaces of constant phase orthogonally. When the
point $P$ is pushed on the boundary, the curves of constant
phase meet, forming a cusp and the locus of such cusps is the caustic. In such a
situation, the standard stationary phase method fails, and one must choose an
appropriate phase function $\Phi$ that parametrizes the Lagrangian manifold locally.

Now, let us return to the reference frame $O(X,Y,Z)$ introduced above in
connection with the proof of Proposition \ref{pro:new6} and introduce ordinary
spherical coordinates $(r,\theta,\varphi)$. Moreover, we introduce geodesic
boundary coordinates as well (defined in a general setting in Subsection \ref{subse:non}),
and now referred to the border of the obstacle. With this in mind, we fix a
point $b$ belonging to the boundary $B$ and denote by $v(r,\theta)$ the
distance from the boundary along the normal to the boundary at $b$ lying in the
plane $XZ$. Recall that the normal coordinate must satisfy a positivity
condition and vanishes on the border. The border of the obstacle is
represented by $u(r,\theta)=0$. Beside these geodesic boundary coordinates, we
also introduce an axis perpendicular to the normal
at $b$ (i.e., tangent to the border at $b$) and lying in the plane $XZ$ and 
denote by $\xi$ the corresponding coordinate. The mapping of the Lagrangian manifold
composed by the ray trajectories generated by the border of the obstacle in
the neighborhood of the point $b$, is of the following form:
$\Lambda(\xi,\cdot)\rightarrow{\mathbb{M}}^*(\xi^2,\cdot)$. In fact, expressing
$v$ as a smooth function of $\xi$ and taking the positivity of $v$ into account,
we are led locally to the equality $v=\xi^2$; moreover, the convexity of the obstacle
(e.g., of a spherical ball) imposes that $u$ depends on $\xi^2$ in a neighborhood of $b$.
The following question arises: Is it possible to determine regular functions
$u$ and $v$ depending on $\xi^2$ only and a phase function $\Phi(u,v;\xi)$
which realizes the mapping $\Lambda(\xi,\cdot)\rightarrow{\mathbb{M}}^*(\xi^2,\cdot)$? 
We answer the previous question with the following proposition.

\begin{proposition}
\label{pro:new7}
The following phase function
\begin{equation}
\Phi(u,v;\xi)=u(r,\theta)+v(r,\theta)\xi-\frac{1}{3}\xi^3
\label{r1}
\end{equation}
gives a local map $\Lambda(\xi,\cdot)\rightarrow{\mathbb{M}}^*(\xi^2,\cdot)$ of
the Lagrangian manifold composed by the ray trajectories generated by the border
of the obstacle in a neighborhood of the point $b$.
\end{proposition}

\noindent
{\bf Proof:} For $\Phi$ of the form (\ref{r1}), the critical set $C$, where
$\partial\Phi/\partial\xi=0$, is given by $v-\xi^2=0$; the positivity condition
of the normal geodesic boundary coordinate is thus satisfied, i.e., $v=\xi^2$.
The equality $\xi=\pm\sqrt{v}$ leads us to separate the two domains: $C^+$
corresponding to $\xi>0$ and $C^-$ corresponding to $\xi<0$. We have
\begin{subequations}
\label{r2}
\begin{eqnarray}
\Phi^+ &\equiv& \Phi(\xi\in C^+) = u+\frac{2}{3}v^{3/2}, \label{r2a} \\
\Phi^- &\equiv& \Phi(\xi\in C^-) = u-\frac{2}{3}v^{3/2}. \label{r2b}
\end{eqnarray}
\end{subequations}
These equalities imply
\begin{subequations}
\label{r3}
\begin{eqnarray}
&&u = \left(\Phi^+ + \Phi^-\right), \label{r3a} \\
&&v^3 = \frac{9}{16}\left(\Phi^+ - \Phi^-\right)^2. \label{r3b}
\end{eqnarray}
\end{subequations}
Let us note that both $(\Phi^+ + \Phi^-)$ and $(\Phi^+ - \Phi^-)^2$ are even
functions of the variable $\xi$. Then we make use of the following auxiliary
Lemma due to Whitney.

\begin{lemma}{{\rm(Whitney \cite{Guillemin})}}
\label{lem:whitney}
Let $f$ be a smooth even function on the real line. Then there exists a smooth
function $g$ on the real line such that $f(x)=g(x^2)$. If $f$ depends smoothly
on a set of parameters, then $g$ can be chosen in such a way that it depends smoothly on the
same parameters\footnote{Editors' note: Certainly, the function $x\to |x|^p, p>1$, $x\in\mathbb{R}$, 
cannot be represented as a smooth function of $x^2$ for $p<2$. Therefore, the function $g$ 
can be nonsmooth at the origin.}.
\end{lemma}
It follows from this auxiliary lemma, since the functions $(\Phi^+ + \Phi^-)$ and
$(\Phi^+ - \Phi^-)^2$ are even, that $u$ and $v^3$ exist as smooth functions
of $\xi^2$\footnote{Editors' note: In the above sense.}. Further we note that, since
$(\partial\Phi/\partial\xi)_{(\xi=0)}=(\partial^2\Phi/\partial\xi^2)_{(\xi=0)}=0$,
while $(\partial^3\Phi/\partial\xi^3)_{(\xi=0)}\neq 0$, the Taylor series for
$(\Phi^+ - \Phi^-)^2$ begins with a nonzero term of order $6$. Thus, $v$ exists,
and it is of order $2$ with respect to $\xi$ \cite{Guillemin}, in agreement
with the positivity condition of the normal geodesic boundary coordinate.
\hfill$\Box$

\vskip 0.2cm

The extension of the stationary phase method suggested by Proposition
\ref{pro:new7} makes us to look for a solution of the wave equation of the
following form:
\begin{equation}
\psi(u,v;k)=\int A(u,v;\xi)e^{{\rm i} k(u+v\xi-\frac{1}{3}\xi^3)}\,d\xi.
\label{r4}
\end{equation}
By the Weierstrass--Malgrange preparation theorem \cite{Malgrange}, we can
find functions $a_0(u,v)$, $a_1(u,v)$, $h(u,v;\xi)$ such that:
\begin{equation}
A(u,v;\xi)=a_0(u,v)+a_1(u,v)\xi+h(u,v;\xi)(v-\xi^2),
\label{r5}
\end{equation}
where $v-\xi^2=\partial\Phi/\partial\xi$. Thus integral (\ref{r4})
can be written in the following way:
\begin{equation}
a_0(u,v)\int e^{{\rm i} k\Phi(u,v;\xi)}\,d\xi+a_1(u,v)\int e^{{\rm i} k\Phi(u,v;\xi)}\xi\,d\xi+
\int h(u,v;\xi)\frac{\partial\Phi}{\partial\xi}e^{{\rm i} k\Phi(u,v;\xi)}\,d\xi.
\label{r6}
\end{equation}
The last term in (\ref{r6}) is
$\int h(u,v;\xi)\frac{1}{{\rm i} k}
\frac{\partial}{\partial\xi}e^{{\rm i} k\Phi(u,v;\xi)}\,d\xi$, which
is of the order of $1/k$. One can then prove that there exist a formal asymptotic
series in $k$, whose first terms are respectively denoted by $A_0$ and $A_1$,
such that
\begin{equation}
\psi(u,v;k)\simeq e^{{\rm i} ku}\left[\frac{A_0}{k^{1/3}}
\int e^{{\rm i}(-k^{2/3}v\xi+\frac{1}{3}\xi^3)}\,d\xi
+\frac{A_1}{{\rm i} k^{2/3}}\int {\rm i}\xi
e^{{\rm i}(-k^{2/3}v\xi+\frac{1}{3}\xi^3)}\,d\xi\right],
\label{r7}
\end{equation}
as $k\rightarrow\infty$. Next, recalling the integral representations of the Airy function and of
its derivative, i.e.,
\begin{subequations}
\label{r8}
\begin{eqnarray}
{\rm Ai}(t) &=& \int e^{{\rm i}(t\xi+\xi^3/3)}\,d\xi, \label{r8a} \\
{\rm Ai'}(t) &=& \int{\rm i}\xi\, e^{{\rm i}(t\xi+\xi^3/3)}\,d\xi,
\label{r8b}
\end{eqnarray}
\end{subequations}
one can finally prove the following proposition (see \cite{Ludwig1}).
\begin{proposition}
\label{pro:new8}
{\rm (i)} In a neighborhood of a fold point of the Lagrangian manifold $\Lambda$
generated at the boundary $\partial{\mathbb{K}}$ of a convex obstacle
${\mathbb{K}}$, the wavefunction $\psi(u,v;k)$ can be represented for high
values of the wavenumber $k$ ($k\rightarrow +\infty$) as follows:
\begin{equation}
\psi(u,v;k)\simeq e^{{\rm i} ku}\left[\frac{A_0}{k^{1/3}}{\rm Ai}(-k^{2/3}v)
+\frac{A_1}{{\rm i} k^{2/3}}{\rm Ai'}(-k^{2/3}v)\right],
\label{r9}
\end{equation}
where $A_0$ and $A_1$ are the first terms of the formal series asymptotic in $k$. \\
{\rm (ii)} The functions $u$ and $v$ satisfy the following system (Ludwig system)
\cite{Ludwig1}:
\begin{subequations}
\label{r10}
\begin{eqnarray}
&&|\nabla u|^2 + v |\nabla v|^2 = 1, \label{r10a} \\
&&\nabla u \cdot \nabla v = 0.\label{r10b}
\end{eqnarray}
\end{subequations}
\end{proposition}

\noindent
{\bf Proof:} (i) The asymptotic representation of $\psi(u,v;k)$ given by
formula (\ref{r9}) follows from formulae (\ref{r4})--(\ref{r8}). \\
(ii) The Ludwig system (\ref{r10}) is obtained by inserting the r.h.s. of
formula (\ref{r9}) into the Helmholtz equation (\ref{2.1}), and
collecting the coefficients suitably.
\hfill$\Box$

~

So far we have treated only those physical problems characterized by the
condition $v(r,\theta)=\xi^2\geqslant 0$. However, there is some
interest in studying phenomena which present the penetration of light
beyond the caustic, i.e., the shadow of the caustic. This physical situation
can be characterized mathematically by setting $v=({\rm i}\eta)^2=-\eta^2$ (i.e.,
$\xi={\rm i}\eta,\,\eta\in{\mathbb{R}}$) in the Ludwig system.
Then the mixed hyperbolic--elliptic character of the Ludwig system emerges.
We study this peculiar feature using polar coordinates $(r,\theta)$.
We prove the following proposition (see also \cite{DeMicheli3}).
\begin{proposition}
\label{pro:new9}
{\rm (i)} The equation of the characteristics for the Ludwig system (\ref{r10}) reads
\begin{equation}
\frac{1}{r^2}\left(u_\theta^2-vv_\theta^2\right)\left(\frac{dr}{d\theta}\right)^2-
2\left(u_ru_\theta-vv_rv_\theta\right)\left(\frac{dr}{d\theta}\right)
+r^2\left(u_r^2-vv_r^2\right)=0.
\label{r11}
\end{equation}
The discriminant $\Delta$ of (\ref{r11}) is
\begin{equation}
\Delta=v\left(u_r v_\theta-u_\theta v_r\right)^2=v |J|^2,
\label{r12}
\end{equation}
where $J$ is the Jacobian determinant $|\frac{\partial(u,v)}{\partial(r,\theta)}|$. \\
{\rm (ii)} The characteristic curves are given by
\begin{equation}
\left(\frac{dr}{rd\theta}\right)=\frac{u_r\pm\sqrt{v}v_r}
{\frac{1}{r}u_\theta\pm\sqrt{v}\frac{1}{r}v_\theta}=
\frac{(\nabla\Phi^\pm)_r}{(\nabla\Phi^\pm)_\theta},
\label{r13}
\end{equation}
where
$\nabla\Phi=(u_r+\sqrt{v}v_r,\frac{1}{r}u_\theta+\sqrt{v}\frac{1}{r}v_\theta)$.
Then
\begin{subequations}
\label{r14}
\begin{eqnarray}
&&\Phi^\pm = u \pm \frac{2}{3}v^{3/2}, \label{r14a} \\
&&|\nabla\Phi^\pm|^2=1. \label{r14b}
\end{eqnarray}
\end{subequations}
{\rm (iii)} According to the sign of $\Delta$ (see \ref{r12}), provided that $|J|\neq 0$, we
can distinguish among the following cases: \\
\indent (a) For $v>0$, $\Delta>0$, we are in the hyperbolic case (the characteristics
are real and distinct); \\
\indent (b) For $v=0$, $\Delta=0$, we are in the parabolic case (the characteristics
are real and coincident); \\
\indent (c) For $v<0$, $\Delta<0$, we are in the elliptic case.
\end{proposition}

\noindent
{\bf Proof:} (i) We rewrite system (\ref{r10}) in terms of the coordinates
$(r,\theta)$, and obtain
\begin{subequations}
\label{r10prime}
\begin{eqnarray}
&&u_r^2+\frac{1}{r^2}u_\theta^2+v\left(v_r^2+\frac{1}{r^2}v_\theta^2\right) = 1,
\label{r10primea} \\
&&u_r v_r + \frac{1}{r^2} u_\theta v_\theta = 0.\label{r10primeb}
\end{eqnarray}
\end{subequations}
The characteristic determinant associated with system (\ref{r10prime}) reads as follows
\begin{equation}
\label{r15}
\left |
\begin{array}{cccc}
2u_r & \frac{2}{r}u_\theta & 2vv_r & \frac{2}{r}vv_\theta \\
dr   & r d\theta   & 0     & 0     \\
v_r  & \frac{1}{r}v_\theta  & u_r   & \frac{1}{r}u_\theta   \\
0    & 0    & dr    & r d\theta
\end{array}
\right |,
\end{equation}
which is equal to
\begin{equation}
\frac{2}{r^2}\left(u_\theta^2-vv_\theta^2\right)dr^2-
4\left(u_ru_\theta-vv_rv_\theta\right)dr d\theta+
2r^2\left(u_r^2-vv_r^2\right)d\theta^2.
\label{r16}
\end{equation}
Then the equation of the characteristics for the Ludwig system is given by
(\ref{r11}). The discriminant $\Delta$ of (\ref{r11}) is
\begin{equation}
\Delta=v(u_r v_\theta-u_\theta v_r)^2 = v|J|^2.
\label{r18}
\end{equation}
(ii) From (\ref{r11}) and (\ref{r18}) it follows that
\begin{equation}
\frac{dr}{d\theta}=r^2\frac{(u_r u_\theta -v v_r v_\theta)
\pm\sqrt{v}(u_r v_\theta-u_\theta v_r)}
{(u_\theta^2-v v_\theta^2)}.
\label{r19}
\end{equation}
The characteristic curves are then given by
\begin{equation}
\left(\frac{dr}{rd\theta}\right)=\frac{u_r\pm\sqrt{v} v_r}
{\frac{1}{r}u_\theta \pm \sqrt{v} \frac{1}{r} v_\theta}\qquad(v>0).
\label{r20}
\end{equation}
Therefore, the constant phase curves are
\begin{equation}
\Phi^\pm = u \pm \frac{2}{3}v^{3/2}\qquad(v>0),
\label{r21}
\end{equation}
which represent branches of cubic curves with negative
($\Phi^+$) and positive ($\Phi^-$) slopes. Let us not that at $v=0$,
these curves have cusps with vertical tangent. Furthermore,
$\nabla\Phi^\pm=\left(u_r\pm\sqrt{v}v_r,\frac{1}{r}u_\theta
\pm\sqrt{v}\frac{1}{r}v_\theta\right)$ satisfties the equation:
\begin{equation}
\left|\nabla\Phi^\pm\right|^2 = 1.
\label{r22}
\end{equation}
(iii) It follows from (\ref{r12}).
\hfill$\Box$

\vspace{2.5ex}

\noindent
{\bf Remark:} In formula (\ref{r9}) the term involving the derivative of the
Airy function is relatively small near the caustic, but significant away from the
caustic \cite{Ludwig1}. This behavior follows from the asymptotic
expansion of the functions ${\rm Ai}$ and ${\rm Ai'}$: the first one is
proportional to $|v|^{-1/4}$, whereas the second one to $|v|^{1/4}$. We shall
return to the use of formula (\ref{r9}) below, in connection with the
specific physical problems in question.

\section{Diffracted rays, orbiting resonances and complex rays}
\label{se:singular}
\subsection{Diffracted rays, singular eigenfunctions, and breaking of quantization}
\label{subse:diffracted}
\subsubsection{Transport equation and damping of the amplitude}
\label{subsubse:transport}
The diffraction problem can be formulated by adding suitable boundary conditions
to the Helmholtz equation (\ref{2.1}): one condition at the border of the
obstacle, and another one at infinity. The first one is an ordinary
boundary condition (e.g., of Dirichlet type); the other is the Sommerfeld
radiation condition. If the symmetry of the diffracting body enables one to apply
the method of separation of the variables, then one can treat the
angular problem separately from the radial one. Furthermore, if the obstacle is
a spherical ball, then the radial problem leads to the first--kind Hankel
functions $H_{n+1/2}^{(1)}(kr)$ (where $n$ is integer, $k$ is the wavenumber,
and $r$ is the distance from the center of the sphere).
The radiation condition is satisfied by the Hankel functions, but we
are obliged to impose, in addition, a boundary condition at the border of
the obstacle. If the latter is of Dirichlet type, then we arrive at the
equation
\begin{equation}
H_{n+1/2}^{(1)}(kR) = 0,
\label{3.1}
\end{equation}
(where $R$ stands for the radius of the spherical ball). The roots of (\ref{3.1}) belong to
the positive imaginary $n$--half--plane, and are infinite in number
\cite{Sommerfeld}. In order to stress that the index of these
functions takes complex values, we replace $n+\frac{1}{2}$ by $\nu_n$.
Accordingly, Eq. (\ref{3.1}) will be rewritten as $H_{\nu_n}^{(1)}(kR)=0$.

\vspace{2.5ex}

\noindent
{\bf Remarks:}
(i) The roots of the equation $H_{\nu_n-1/2}^{(1)}(kR)=0$ are given by the rule \cite{Sommerfeld}
\begin{equation}
\nu_n = kR + \frac{1}{2}(kR)^{1/2}\left[\frac{3}{4}\pi(4m-1)\right]^{2/3}
e^{{\rm i}\pi/3},
\label{n2}
\end{equation}
$(m=1,2,3,\ldots)$, and at any fixed value of $kR$, $\nu_n$ varies by varying $m$.
They are located close to a curve which tends to become parallel to the imaginary
axis of the $\nu$--plane. \\
(ii) It has been pointed out by Sommerfeld \cite{Sommerfeld} that the functions
$H_{\nu_n}^{(1)}(kr)$ form an orthogonal system. If we replace $kr$ by a real
variable $x$, and assume that $k=1$, we have:
\begin{equation}
\int_R^{+\infty} H_{\nu_n}^{(1)}(x) H_{\nu_m}^{(1)}(x)\,\frac{dx}{x} = 0
\qquad (m\neq n).
\label{3.2}
\end{equation}
This orthogonality relation is of an unconventional kind since the Hankel
functions $H_{\nu_n}^{(1)}(x)$ are complex--valued, but no conjugation occurs in
(\ref{3.2}). One could be tempted to reconsider the problem of the waves
propagating along the spherical surface regarding the Hankel functions
$H_{\nu_n}^{(1)}$ as eigenfunctions of a non--selfadjoint boundary value problem
for Bessel's differential operators. But, unfortunately, the Hankel functions
$H_{\nu_n}^{(1)}(x)$ do not form a complete set of functions
\cite{Sommerfeld,Pflumm}.

\vspace{2.5ex}

A solution of the diffraction problem can be written as a formal series. In
this expansion, the angular dependence is represented by Legendre
polynomials (in the case of spherical diffracting bodies),
the radial dependence by Hankel functions of the first kind, and the
coefficients present at the denominator the terms $H_{\nu_n}^{(1)}(kR)$.
However, these series converge very slowly and are, in fact, useless for numerical
computation. Following Sommerfeld \cite{Sommerfeld}, one
can perform a Watson--type resummation of these expansions, transforming a
sum over $n$ ($n$ is an integer) into an integral along a suitable path in the
complex $\nu$--plane \cite{Sommerfeld}. Then the poles of the integrand
under consideration are the roots of the equation $H_{\nu_n}^{(1)}(kR)=0$.
The sum over the residues at the poles located in the first quadrant of the
$\nu$--plane is rapidly convergent for values of the angle sufficiently large
(i.e., backwards). In particular, the term corresponding to the pole closest
to the real axis is the dominant one, and, in general, it is sufficient for
describing the diffraction in the backward angular region.

Coming back to the geometrical approximation, and specifically to the CFUL
approximation formula (\ref{r9}), we note, first of all, that the term
$v(r,\theta)$ is zero at $\xi=0$ (i.e., on the border of the
obstacle). It follows that formula (\ref{r9}) cannot satisfy the boundary
condition (e.g., the Dirichlet condition) at the surface of the obstacle. Lewis,
Bleinstein and Ludwig \cite{Lewis} proposed a new {\it ansatz}, which is a
slight modification of formula (\ref{r9}), in order to
obtain an approximation suitable for imposing appropriate boundary conditions.
We follow a simpler approach based on some assumptions which can be controlled
from the physical viewpoint. First of all, we choose Ludwig's system
(see (\ref{r10})), which describes the geometry of rays
in a neighborhood of the caustic generated by the border of the obstacle.
Next, we note that the poles (in the $\nu$--plane) generated by
the roots of the equation $H_{\nu_n}^{(1)}(kR)=0$ (which follow by imposing
the Dirichlet boundary conditions) describe the damping of the density
of the flux of the trajectories bending around the obstacle. It follows that
it is precisely the transport equation that should be appropriately modified.
With this in mind, we assume that the leakage of rays from the tube of
trajectories running along the boundary of the obstacle is proportional
to the flux. If we suppose that the diffracting body is a spherical ball,
then the transport equation on the border of the obstacle reads as follows
(see also \cite{Levy}):
\begin{equation}
\frac{1}{|J_0|}\frac{d}{d\tau}\left(A_0^2 |J_0|\right)=-2\gamma_0 A_0^2,
\label {3.5}
\end{equation}
where $J_0$ is the Jacobian
\begin{equation}
J_0=\left[\frac{\partial(x,y,z)}{\partial(r,\varphi,\tau)}
\right]_{\staccrel{\!\!\!\!\!\!\scriptstyle r=R}
{\scriptstyle\tau^\pm=R\theta_0^\pm}},
\label{t1}
\end{equation}
and $\gamma_0$ depends upon the curvature of the obstacle and is constant
in the case of a diffracting spherical ball. It follows from (\ref{3.5}) and
(\ref{t1}) for $\theta_0^\pm\neq (n+\frac{1}{2})\pi$ ($n=0,1,2,\ldots$) that
\begin{equation}
A_0(\theta_0^\pm)=\frac{C}{\sqrt{|\cos\theta_0^\pm|}}\,e^{-\gamma_0 R\theta_0^\pm}
\qquad (C={\mbox{Const.}}).
\label{3.6}
\end{equation}
The meaning of $\gamma_0$ can be illustrated by returning to Proposition
\ref{pro:3}. It follows from statement (b) of Proposition \ref{pro:3} that,
at any point of the boundary, the ray splits into two rays: one part travels along the
surface as a geodesic bending around the obstacle whereas the other leaves the
diffracting body tangentially. Then $\gamma_0$, at any chosen value of the wavenumber $k$,
depends on the curvature of the boundary, and this is true not only
when the obstacle is a spherical ball but also for any compact and convex diffracting
object. Finally, the zero subscript in the notation of $\gamma_0$ is for
recalling that this damping factor corresponds to the mode associated with the
pole closest to the real axis in the Sommerfeld--type
summation illustrated above in the particular case of obstacles represented
by spherical balls embedded in ${\mathbb{R}}^3$ (recall that only this type of 
diffracting bodies will be considered in the following).

\subsubsection{Solution of the Ludwig system at $\mathit{v>0}$}
\label{subsubse:solution}
First, let us consider the counterclockwise oriented rays. We come back to the
Ludwig system written in terms of coordinates $(r^+,\theta_0^+)$, appropriate
to the counterclockwise trajectories. Notice that the orientation
of $r^+$ coincides with that of the standard radial coordinate, and
$\theta_0^+$ was introduced in Subection \ref{subse:geometric}.
We now write $u(r^+,\theta_0^+)=\theta_0^+$ (assuming $R=1$, for simplicity); 
accordingly, $u_{r^+}=0$, $u_{\theta_0^+}=1$. It follows from (\ref{r10primeb})
that $v_{\theta_0^+}=0$.  Moreover, from (\ref{r10primea}) we get, for large
values of $r^+$:
\begin{equation}
\sqrt{v}\,dv = \sqrt{1-\left(\frac{1}{r^+}\right)^2}\,dr^+
\staccrel{\sim}{r^+\rightarrow +\infty} dr^+.
\label{p1}
\end{equation}
Integrating, we obtain
\begin{equation}
\frac{2}{3}v^{3/2} \staccrel{\sim}{r^+\rightarrow +\infty} r^+.
\label{p2}
\end{equation}
Next, we return to the expression of $\Phi^+$, and see that
\begin{equation}
\Phi^+ = u + \frac{2}{3}v^{3/2} = \theta_0^+ + r^+.
\label{p3}
\end{equation}
Analogous calculations can be repeated for the clockwise rays. Taking
$u(r^-,\theta_0^-)=\theta_0^-$, we have
\begin{equation}
\Phi^- = u - \frac{2}{3}v^{3/2} = \theta_0^- - r^-.
\label{p4}
\end{equation}
Now, we relate the coordinates $(r^\pm,\theta_0^\pm)$ to the coordinates
appropriate for describing the outgoing rays at large distance from the obstacle, i.e.,
the scattering coordinates: $r_s\equiv r$ and $\theta_s$ ($0<\theta_s<\pi$).
We have: $\theta_0^+=\theta_s$, $r^+=r_s\equiv r$;
$\theta_0^-=2\pi-\theta_s$, $r^-=-r_s=-r$. Therefore, in conclusion we have:
\begin{subequations}
\label{p5}
\begin{eqnarray}
\Phi^+ &=& \theta_s + r, \label{p5a} \\
\Phi^- &=& 2\pi-\theta_s + r. \label{p5b}
\end{eqnarray}
\end{subequations}
So far we have considered trajectories which have not completed a single turn. However
the ray trajectories can perform several turns around the obstacle before
emerging, and accordingly, the flux density along the border of the obstacle
is not conserved, as we have shown above. \emph{It follows
that the wavefunction does not return to the original value after one turn, and
it is not single--valued.} We are thus obliged to sum the contributions of
both the counterclockwise and clockwise trajectories turning around the obstacle
several times, and then emerging and contributing to the scattering amplitude.

\subsubsection{Eikonal approximation and its uniformization}
\label{subsubse:eikonal}
We now have to come back to formula (\ref{2.24}) and to the analogous one for
the clockwise rays. First we note that
\beq
\tau^\pm=\sqrt{(r^\pm)^2-R^2}\staccrel{\longrightarrow}{r\rightarrow+\infty} r,
\eeq
and then
\beq
\sqrt{|J|}\staccrel{\longrightarrow}{r\rightarrow+\infty}r
\sqrt{|\sin\theta_0^\pm|}=r\sqrt{\sin\theta_s}.
\eeq
We are thus led to consider the branch--cut singularities associated
with the crossing of the trajectories through the axial caustic once again. As we have
already noted, $\theta_0^+=\theta_s$ for the
counterclockwise rays which have not completed one turn, and more generally,
$\theta_{0,n}^+-2\pi n=\theta_s$ $(n=0,1,2,\ldots)$ for the rays which have
performed $n$ turns around the sphere before emerging. But the axial caustic
corresponds to $\theta_s=0$ or $\pi$; accordingly,
$\theta_{0,n}^+=2\pi n$ or $(2n+1)\pi$.
Analogous considerations hold for the clockwise rays.
Therefore, proceeding as in the proof of Proposition \ref{pro:2} (see also
Proposition \ref{pro:5}), we can
associate a phase--shift to each crossing of the ray trajectories through
the axial caustic (including the antipodal points of the spherical ball).
More precisely:
\begin{itemize}
\item[(i)] at each counterclockwise crossing of the axial caustic we have a
phase--shift of $e^{-{\rm i}\pi/2}$;
\item[(ii)] at each clockwise crossing of the axial caustic we have a
phase--shift of $e^{+{\rm i}\pi/2}$.
\end{itemize}
Therefore, the bijective correspondence between the winding number $w(\gamma)$
and the crossing number $n(\gamma)$ (see statement (v) of Proposition \ref{pro:5})
can be extended to the phase--shift associated with the crossing number.
Therefore we have:
\begin{center}
\begin{tabular}{cccccc}
${\mathbb{Z}}$ & $\longrightarrow$ & ${\mathbb{N}}$ & $\longrightarrow$
& {\rm Crossing phase--shift} & \\[+5pt]
$n$  &                   & $(2n+1)$ & $\longrightarrow$ & $-\frac{\pi}{2}(2n+1)$&
$(n \geqslant 0)$ \\[+5pt]
$n$  &                   & $-2n$ & $\longrightarrow$ & $\frac{\pi}{2}(-2n)$ &
$(n < 0)$
\end{tabular}
\end{center}
In view of these formulae and of the analysis performed above, we may work out a
uniformized eikonal approximation. More precisely, the wavefunction
$\psi_{(0)}^{(+)}(r,\theta_s;k)$ associated with a counterclockwise ray which
has not completed one turn presents the following asymptotic behavior for large
values of $r$ and $k$:
\begin{equation}
\psi_{(0)}^{(+)}(r,\theta_s;k)
\staccrel{\longrightarrow}{\staccrel{r\rightarrow\infty}
{\scriptstyle k\rightarrow\infty}}
C(k)\frac{e^{-{\rm i}\pi/2}\,e^{{\rm i} k(R\theta_s+r)}e^{-\gamma_0 R \theta_s}}
{r\sqrt{\sin\theta_s}}\qquad(0<\theta_s<\pi),
\label{p7}
\end{equation}
where $C(k)$ is a diffraction coefficient depending only on $k$, the term
$e^{-{\rm i}\pi/2}$ corresponds to a single crossing of the ray across the axial
caustic, $e^{-\gamma_0 R \theta_s}$ is the damping due
to the leakage of the tube of trajectories along the border of the obstacle
which is supposed to be a spherical ball of radius $R$ (i.e., we then write
$R\theta_s$ instead of $\theta_s$).
The scattering amplitude turns out to be given by:
\begin{equation}
f_{(0)}^{(+)}(\theta_s;k)=C(k)\frac{e^{-{\rm i}\pi/2}\,e^{{\rm i}\nu_0\theta_s}}
{\sqrt{\sin\theta_s}}
\label{p8}
\end{equation}
$(0<\theta_s<\pi, \nu_0=R(k+{\rm i}\gamma_0))$. Note that the subscript zero in
$\psi_{(0)}^{(+)}$ and in $f_{(0)}^{(+)}$ indicates that we are referring to
the ray which has not completed one turn, while
the subscript in $\nu_0$ is for recalling that $R(k+{\rm i}\gamma_0)$ is a
geometrical approximation to the first root of $H_{\nu_n}(kR)=0$: i.e.,
the root closest to the real axis.

We can similarly evaluate the contribution to the scattering amplitude
of the diffracted ray which travelling along the sphere, in clockwise direction,
without completing one turn around the obstacle. We obtain
\begin{equation}
f_{(0)}^{(-)}(\theta_s;k)=(-1)C(k)\frac{e^{{\rm i}
\nu_0(2\pi-\theta_s)}}{\sqrt{\sin\theta_s}}\qquad(0<\theta_s<\pi),
\label{p9}
\end{equation}
where the factor $(-1)$ is precisely given by the product of two phase--shifts,
where each of them is expressed by the exponential $e^{{\rm i}\pi/2}$ since this
grazing ray crosses the axial caustic twice.

\vspace{2.5ex}

\noindent
{\bf Remark:} In the derivation of formulae (\ref{p8}) and (\ref{p9}), instead of
using formula (\ref{r9}) (i.e., the CFUL approximation), we have used the eikonal
approximation implemented by the uniformization procedures derived in Section
\ref{se:riemannian}. We have followed this method
since the Dirichlet boundary condition cannot be applied to formula (\ref{r9}),
as we have already observed before. Further, the ray--tracing procedure, which we
have implemented, is more satisfactory from the physical viewpoint and very close
to Feynman's path method. We recall, finally, that formula (\ref{r9})
implies again the eikonal approximation by using the asymptotic
behavior of ${\rm Ai}$ and ${\rm Ai'}$ for large $v$ and $k\rightarrow\infty$
\cite{Ludwig1}.

\subsubsection{Construction of singular eigenfunctions and breaking of quantization}
\label{subsubse:construction}
Adding $f_0^{(+)}$ and $f_0^{(-)}$, we obtain the scattering amplitude $f_{0}$ due
to the rays not completing one turn around the obstacle:
\begin{equation}
f_{0}(\theta_s;k)=f_0^{(+)}+f_0^{(-)}=
-{\rm i} C(k)\frac{e^{{\rm i}\nu_0\theta_s}-{\rm i}
e^{{\rm i}\nu_0(2\pi-\theta_s)}}{\sqrt{\sin\theta_s}},
\label{3.10}
\end{equation}
$(0<\theta_s<\pi)$. Now, we are ready to take into account the contributions of
all those rays that are orbiting around the sphere several times: consider
rays performing $n$ ($n\in{\mathbb{N}}$) turns around the obstacle.
Since the surface angles $\theta_{0,n}^{(\pm)}$ are related to the scattering
angle $\theta_s$ by the rule $\theta_{0,n}^{(+)}=\theta_s+2\pi n$,
$\theta_{0,n}^{(-)}=2\pi-\theta_s+2\pi n$ ($n=0,1,2,\ldots$), we have, for
$0<\theta_s<\pi$:
\begin{equation}
f(\theta_s;k)=-{\rm i} C(k)\sum_{n=0}^\infty (-1)^n e^{{\rm i} 2\pi n\nu_0}
\frac{e^{{\rm i}\nu_0\theta_s}-{\rm i} e^{{\rm i}\nu_0(2\pi-\theta_s)}}
{\sqrt{\sin\theta_s}}.
\label{3.11}
\end{equation}
We note that the factor $(-1)$, for any $n>0$, is due to the product of two
phase--shifts represented by the exponential factor $(e^{-{\rm i}\pi/2})^2$
for the counterclockwise oriented rays, and by $(e^{{\rm i}\pi/2})^2$ for the clockwise
trajectories. In fact, both the counterclockwise and the clockwise
oriented rays cross the axial caustic twice for each turn (see Fig. \ref{fig_2}).

Now, by exploiting the following expansion:
\begin{equation}
\frac{1}{2\cos\pi\nu_0}=e^{{\rm i}\pi\nu_0}\sum_{n=0}^\infty (-1)^n
e^{{\rm i} 2\pi n\nu_0}
\qquad ({\mbox{Im}\,}\nu_0>0),
\label{3.12}
\end{equation}
one can rewrite the amplitude (\ref{3.11}) as
\begin{equation}
f(\theta_s;k)=-C(k) e^{{\rm i}\pi/4}
\frac{e^{-{\rm i}[\nu_0(\pi-\theta_s)-\pi/4]}+e^{{\rm i}
[\nu_0(\pi-\theta_s)-\pi/4]}}
{2\cos\pi\nu_0\sqrt{\sin\theta_s}},
\label{3.13}
\end{equation}
$(0<\theta_s<\pi)$. The r.h.s. of (\ref{3.13}) gives the asymptotic behavior
of the Legendre function $P_{\nu_0-1/2}(-\cos\theta_s)$ times
$\sqrt{2\pi\nu_0}$ for $|\nu_0|\rightarrow\infty$ and $|\nu_0|(\pi-\theta_s) \gg 1$ \cite{Erdelyi}. 
Then, writing $P_{\nu_0-1/2}(-\cos\theta_s)$ instead
of its asymptotic behavior, we have
\begin{equation}
f(\theta_s;k) \simeq G(k) \frac{P_\lambda(-\cos\theta_s)}{\sin\pi\lambda},
\label{3.14}
\end{equation}
as $|\nu_0|\rightarrow\infty$ and for $0<\theta_s<\pi$,
where $G(k)=C(k)e^{{\rm i}\pi/4}\frac{\sqrt{\pi}}{2}\sqrt{2\lambda+1}$,
$\lambda=\nu_0-\frac{1}{2}$. We have thus obtained the expression of the
scattering amplitude in terms of Legendre functions, which represent
the angular part of the so--called {\it singular eigenfunctions} \cite{Sommerfeld}.
Let us in fact note that $P_\lambda(-\cos\theta_s)$ presents a
logarithmic singularity at $\theta_s=0$ \cite{Sommerfeld}, which indicates
that the representation (\ref{3.14}) of the scattering amplitude fails in the forward
direction. Indeed, at small angles, the surface waves associated with the diffracted
rays describe a very small arc of the circumference, and the damping factors
$\exp(-\gamma_0 R\theta_0^{(\pm)})$ (see (\ref{3.6})) are very close to one.
We are then forced to consider the entire set of modes
corresponding to the countable set of the roots of the equation
$H^{(1)}_{\nu_n}(kR)=0$. Conversely, at large angles, the root
closest to the real axis gives the main contribution, and this is in
agreement with the fact that the factor $\gamma_0 R\theta_0^{(\pm)}$ is large
backwards. Accordingly, $P_\lambda(-\cos\theta_s)=1$ at $\theta_s=\pi$.
Let us then note that ${\mbox{Re}\,}\nu_0=kR={\mbox{Re}\,}\lambda+\frac{1}{2}=
\ell+\frac{1}{2}$ ($\ell$ integer), in agreement with the semiclassical
formula for the angular momentum. Therefore the function ${\mbox{Re}\,}\lambda(k)$
($k$ being the wavenumber) gives the trajectory in the complex $\lambda$--plane
of the scattering amplitude singularity closest to
the real axis. Coming back to formula (\ref{3.14}), let us however note that
if the term $R\gamma_0$ is large, then this pole does not produce sharp peaks
in the observable quantity (i.e., in the scattering
cross--section $\sigma(\theta_s;k)=|f(\theta_s;k)|^2$), in view of the factor
$|\sin\pi\lambda|^{-1}\simeq \exp(-\pi\gamma_0R)$.

As we have seen before (see formula (\ref{2.19})), the geometrical quantization
implies that the angular momentum is given by
$\ell+\frac{1}{2}$ $(\ell=0,1,2,\ldots)$. Now, the leakage of rays from the tube
of trajectories running along the boundary of the obstacle gives rise to a damping
factor, which breaks the quantization rule. The angular wavefunction is, in
diffractive processes, represented by the Legendre function
$P_\lambda$ $(\lambda\in{\mathbb{C}})$ rather than by the Legendre polynomials
$P_\ell$ $(\ell=0,1,2,\ldots)$, and the angular symmetry is broken.
For a detailed phenomenological analysis of the creeping waves in the
$\pi^+$--p elastic scattering, see \cite{DeMicheli4}.

\subsection{Orbiting resonances and bound states}
\label{subse:orbiting}
The orbiting resonances observed in molecular scattering are produced by the
trapping of the incoming colliding particles in the region
delimited by the centrifugal barrier on one side and by the hard--core on the
other side. The process can be depicted as follows: the trajectories of the
incoming particles hit the impenetrable sphere which
composes the hard--core and describe geodesic orbits around it. This part of
the process is quite similar to diffraction by a totally opaque sphere outlined
in the previous subsection, and it can be described by using the same geometrical
tools. We can thus speak of a tube composed by those trajectories
that describe circular orbits bending the spherical ball that represents the
hard--core. However, now the leakage of rays from the tube composed by orbiting
trajectories is attenuated by the centrifugal barrier. Therefore we shall find
asymptotically (i.e., at large $r$) only the particles crossing the
centrifugal barrier by tunnelling. At this point, the following problem emerges:
in the tunnelling the energy $E$ of the particle is less than the height $V$ of
the potential barrier; accordingly, $n^2=1-\frac{V}{E}$ is negative, and
therefore it cannot be identified with a Riemannian
metric tensor. We loose the identification of a trajectory
with a Riemannian real--valued geodesic. We can, however, extend the concept of
ray by admitting complex--valued and/or imaginary phases, and speak of complex
trajectories in the sense of Landau \cite{Landau}.
Coming back to the standard eikonal equation, an imaginary refractive index
produces an imaginary phase. Therefore, the flux of particles beyond the
centrifugal barrier is attenuated by a damping factor which is produced by
the imaginary phase, and it is proportional to the ratio between the height of
the barrier and the energy of the particles. The lifetime of the resonances is
linked to the leakage of the tube of orbiting trajectories,
and is controlled by the tunnelling across the centrifugal barrier. Coming back
to the calculations of the previous subsection, we simply replace the
factor $R\gamma_0$ a factor denoted by $\beta$ (the subscript zero is now omitted
for simplicity), which is related to the lifetime of the resonance: the longer
is the life of the resonance the smaller is the value of $\beta$. At $\beta=0$,
the lifetime is infinite, the state is stable: we have orbiting bound states.

Let us then return to formula (\ref{3.14}), with the notation
$\lambda=\alpha+{\rm i}\beta$. If $\alpha=\ell$
($\ell$ is an integer) and $0<\beta\ll 1$, then the factor $1/\sin\pi\lambda$ produces a
sharp peak in the cross--section: we have a resonance in the $\ell$--th partial
wave. By the formula
\begin{equation}
\frac{1}{2}\int_{-1}^{+1}P_\lambda(-z)P_\ell(z)\,dz=
\frac{\sin\pi\lambda}{\pi(\lambda-\ell)(\lambda+\ell+1)}
\qquad (\ell=0,1,2,\ldots;\lambda\in{\mathbb{C}}),
\label{PP}
\end{equation}
we can project the amplitude given by the r.h.s. of (\ref{3.14}) on the
$\ell^{\rm th}$ partial wave
and (by using the standard notation) obtain the following expression
(see \cite{DeMicheli2}):
\begin{equation}
a_\ell=\frac{e^{2{\rm i}\delta_\ell}-1}{2{\rm i} k} =
\frac{C(E)}{\pi}\frac{1}{(\alpha+{\rm i}\beta-\ell)(\alpha+{\rm i}\beta+\ell+1)},
\label{n3}
\end{equation}
($\lambda=\alpha+{\rm i}\beta$, $E=$ center of mass energy, $k^2=E$ in suitable
units). Next, when the elastic unitarity condition can be used,
we obtain \cite{DeMicheli2}
\begin{equation}
C(E)=-\frac{\pi}{k}\beta(2\alpha+1),
\label{EE}
\end{equation}
and then
\begin{equation}
\delta_\ell=\sin^{-1}
\frac{\beta(2\alpha+1)}{\{[(\ell-\alpha)^2+\beta^2]
[(\ell+\alpha+1)^2+\beta^2]\}^{1/2}},
\label{n4}
\end{equation}
which represents a sequence of orbiting resonances, i.e., an ordered set of
phase--shifts $\delta_\ell$ that cross $\frac{\pi}{2}$ with positive derivative.
In the present analysis, we do not consider the downward
crossing of the phase--shifts through $\frac{\pi}{2}$, i.e., the echoes of the
resonances (or antiresonances);
the interested reader is referred to \cite{DeMicheli2}.
Now, the l.h.s. of (\ref{n1}) (see Section \ref{se:geometrical})
can be extended to continuous values of the angular momentum $\ell$ by writing
\begin{equation}
\alpha(\alpha+1)=2IE + c_0.
\label{n5}
\end{equation}
As we already remarked in Section \ref{se:geometrical}, it is precisely the
extrapolation of this straight line from $E>0$ to $E<0$ that allows us to
interpolate the orbiting bound states at $E<0$.

\subsection{Complex rays. Evanescent waves in the shadow of the caustic:
The rainbow}
\label{subse:evanescent}
\subsubsection{Generalized Cauchy--Riemann equations}
\label{subsubse:generalized}
We now consider the case $v<0$, i.e., statement (iii) of Proposition \ref{pro:new9}.
The Ludwig system is elliptic. Equations (\ref{r10}) inform us that the
gradients of $u$ and $v$ are orthogonal; then, where $|J|\neq 0$, this amounts
to writing
\begin{equation}
\label{s1}
\frac{
\left[
\begin{array}{c}
u_r \\
\frac{1}{r}u_\theta
\end{array}
\right]
}{|\nabla u|}
=
\frac{
\left[
\begin{array}{c}
\frac{1}{r} v_\theta \\
-v_r
\end{array}
\right]
}{|\nabla v|},
\end{equation}
from which the equalities
\begin{subequations}
\label{s2}
\begin{eqnarray}
u_r &=& \rho \, \frac{1}{r}\, v_\theta, \label{s2a} \\
u_\theta &=& -\rho \, r \, v_r, \label{s2b}
\end{eqnarray}
\end{subequations}
follow, with $\rho(r,\theta)=|\nabla u|/|\nabla v| =
(\frac{1}{|\nabla v|^2}-v)^{1/2} > 0$, since $v<0$.
Equations (\ref{s2}) are a generalization of the Cauchy--Riemann equations
written in polar coordinates.

Coming back to the expression of the phase $\Phi^\pm$,
we have: $\Phi^\pm=u\mp{\rm i}\frac{2}{3}(-v)^{3/2}$,
namely, a complex--valued phase naturally emerges as a consequence of
the elliptic character of the eikonal system (Ludwig system) for $v<0$.

\subsubsection{Solution of the Ludwig equations at $v<0$, and the evanescent waves}
\label{subsubse:solution2}
Writing again $u(r^\pm,\theta_0^\pm)=\theta_0^\pm$ in (\ref{r10}) and
(\ref{r10prime}) and supposing once again that $R=1$, we obtain: $u_{r^\pm}=0$,
$u_{\theta_0^\pm}=1$, $v_{\theta_0^\pm}=0$. Specifically, from (\ref{r10primea}) we have
\begin{equation}
\frac{1}{r^2}-1=(-v)v_r^2 \qquad (v<0).
\label{s3}
\end{equation}
Since we want to describe the shadow of the circular caustic, we integrate
(\ref{s3}) over the domain inside the unit circle, i.e., over the interval $(r,1)$,
$0<r<1$. We thus obtain, for $0<r<1$:
\begin{equation}
-\frac{2}{3}(-v)^{3/2}=\int_1^r\sqrt{\frac{1}{{r'}^2}-1}\,d{r'}=
\sqrt{1-r^2}-\frac{1}{2}\ln\frac{1+\sqrt{1-r^2}}{1-\sqrt{1-r^2}}.
\label{s4}
\end{equation}
Since the approximation of the Airy function ${\rm Ai}(-k^{2/3}v)$ (see formula
(\ref{r9})) is given, in the domain $v<0$, by the following asymptotic formula
\begin{equation}
{\rm Ai}(-k^{2/3}v)\simeq
\frac{1}{2\sqrt{\pi}k^{1/6}}\frac{e^{-\frac{2}{3}k(-v)^{3/2}}}{(-v)^{1/4}},
\label{s5}
\end{equation}
we have from formulae (\ref{s4}) and (\ref{s5}) that
\begin{equation}
{\rm Ai}(-k^{2/3}v)\simeq\frac{g(r)}{2\sqrt{\pi}k^{1/6}}
\exp\left(k\sqrt{1-r^2}-\frac{k}{2}\ln\frac{1+\sqrt{1-r^2}}{1-\sqrt{1-r^2}}\right).
\label{s6}
\end{equation}
Now, let us note that, in the approximation (\ref{r9}) (i.e., the CFUL approximation),
we can retain the first term only if we are in a region close to the caustic.
The second term, proportional to ${\rm Ai'}$, is relevant only far from the caustic.
But, in any case, it can be neglected in view
of its exponentially decreasing behavior. We thus obtain
\begin{equation}
\psi_S(r,\theta_0^\pm;k)\simeq
\frac{G_S(r,\theta_0^\pm)}{\sqrt{k}}e^{{\rm i} k\theta_0^\pm}\,
\exp\left[k\left(\sqrt{1-r^2}-\frac{1}{2}
\ln\frac{1+\sqrt{1-r^2}}{1-\sqrt{1-r^2}}\right)\right],
\label{s7}
\end{equation}
where the subscript $S$ is to indicate that we treat the shadow region.
When setting $y^2=1-r^2$, formula (\ref{s7}) can be rewritten as follows:
\begin{eqnarray}
\lefteqn{\psi_S(y,\theta_0^\pm;k)\simeq
\frac{G_S(y,\theta_0^\pm)}{\sqrt{k}}e^{{\rm i} k\theta_0^\pm}
e^{ky}\,e^{-\frac{k}{2}\ln\frac{1+y}{1-y}}} \nonumber \\
&&= \frac{G_S(y,\theta_0^\pm)}{\sqrt{k}}e^{{\rm i} k\theta_0^\pm}
\left(\frac{1-y}{1+y}\right)^{k/2}\,e^{ky}
\qquad (y^2=1-r^2, 0\leqslant r\leqslant 1). \label{s8}
\end{eqnarray}
On the circular caustic, i.e., $r=1$, we have $y=0$ and
$\left(\frac{1-y}{1+y}\right)^{k/2}\,e^{ky}=1$; then
from (\ref{s8}), we remain with only counterclockwise $e^{{\rm i} k\theta_0^+}$
or clockwise $e^{{\rm i} k\theta_0^-}$ waves propagating on the unit circle.
For $r\rightarrow 0$,
we have $y^2\rightarrow 1$, and the only physical admissible solution is for
$y=+1$. Then, $\left(\frac{1-y}{1+y}\right)^{k/2}\,
e^{ky}\staccrel{\longrightarrow}{y\rightarrow 1}0$,
which describes the damping of the evanescent wave in the shadow of the caustic.

In order to study the shadow of the rainbow, it is convenient to map the interior
of the unit disk into the outside part, by setting $r'=1/r$, and accordingly,
$y'^2=1-1/r'^2$. Therefore, formula (\ref{s8}) reads (for simplicity, we omit the prime):
\begin{equation}
\psi_S(y,\theta_0^\pm;k)
\simeq\frac{G_S(y,\theta_0^\pm)}{\sqrt{k}}e^{{\rm i}
k\theta_0^\pm}\left(\frac{1-y}{1+y}\right)^{k/2}e^{ky}
~~~~(y^2=1-\frac{1}{r^2}; 1\leqslant r < +\infty).
\label{s9}
\end{equation}
Then for $r=1$ we have $y=0$ and $\left(\frac{1-y}{1+y}\right)^{k/2}e^{ky}=1$;
for $r\rightarrow+\infty$, $y\rightarrow 1$, and then
$\left(\frac{1-y}{1+y}\right)^{k/2}e^{ky}\rightarrow 0$,
which gives the damping of the evanescent waves in
the shadow of the caustic.

The rainbow exemplifies the simplest caustic where two rays, deflected by
refraction and reflection in a raindrop, coalesce. The two rays coalesce at the
rainbow angle, and on the dark side there are no real rays.
As examples of rainbow, we can consider the nuclear (on which we focus
particularly our attention) and the Coulomb rainbows, which can be found in the
scattering phenomenology, e.g., in $\alpha$--$^{40}$Ca elastic scattering
\cite{Delbar}. The rainbow scattering is a case where the classical cross--section
has a shadow region. For scattering angle $\theta_s$ less than the critical
angle $\theta_r$ (the rainbow angle), two classical
trajectories contribute to the cross--section. On the other hand, no classical
real--valued orbit has a scattering angle $\theta_s>\theta_r$, and there is a shadow.
In a neighborhood of the rainbow angle $\theta_s\simeq\theta_r$, there is
a bump in the cross--section; beyond $\theta_r$, the cross--section
decreases almost exponentially, and this behavior is in agreement with formula
(\ref{s9}) for sufficiently high values of $k$. Finally, it is worth mentioning
the spectacular example of rainbow dip in the differential
cross--section of $\alpha$--$^{40}$Ca, i.e., the dip at the energy
of $48.0$ MeV (in the laboratory system), which can be associated with the
nuclear rainbow generated by the refracted rays which penetrate the interaction
region, without being absorbed, and then at the rainbow angle
$\theta_r\simeq 120^\circ$ (deg. c.m.) coalesce (see Ref. \cite{Delbar}).
Let us note that this phenomenon of rainbow is closely connected with a
remarkable aspect of the $\alpha$--nucleus scattering, i.e.,
{\it the reduced absorption of the particles penetrating the target nucleus}.
This is caused by the large binding energy of the nucleons in $\,^4$He and the
saturation of the nucleon--nucleon forces in cases (like the $\alpha$--$^{40}$Ca)
of projectiles and target nuclei with closed shells \cite{VonOertzen}.

\section{Conclusions}
\label{se:conclusions}
As is well--known, the solution of the wave equation obtained by the method
of the geometrical optics is related, in a sense, to the asymptotic form
of the integral representation of the field (if it exists), which is an exact
solution of the wave problem. Suppose, for example, that the field, in a uniform medium,
can be written in the form of an expansion in plane waves; the evaluation of this integral
by the stationary phase method yields an asymptotic series. One then extracts the leading
term of this asymptotic expansion, which is composed by an amplitude and a phase.
The ray trajectories are the lines orthogonal to the constant phase surface, and
are described by the eikonal equation; the amplitude satisfies the transport equation,
whose physical meaning is the conservation of the flux density. In the simplest
case of uniform medium, whose refractive index $n$ is a real constant, the rays
are straight lines which can be characterized by the following properties:\\
\indent (i) They are geodesics of the Euclidean space. \\
\indent (ii) Phase and amplitude are real--valued functions. \\
\indent (iii) They can be derived by the Fermat's variational principle.

Constrained by these types of properties the methods of geometrical optics are rather limited and
fail to explain several phenomena like, for instance, the diffraction by
a compact and opaque obstacle, that is, the existence of non--null field in the
geometrical shadow which, for this reason, is usually referred to as the classically (or
geometrically) forbidden region.

In the present paper we have tried, following the ideas and methods proper of GTD,
to widen the area of application of geometrical optics. We adopt, first of all,
the Jacobi form of the principle of least action (instead of the Fermat's),
which is concerned with the path of the system point rather than with its time
evolution \cite{Goldstein}. More precisely, Jacobi's principle (generally
applied in mechanics) can be formulated as follows: If there are no forces
acting on the body, then the system point travels along the shortest path
length in the configuration space. Here we assume a wide extension of
Jacobi's principle, which can be formulated as follows: The geodesics
associated with the Riemannian metric (written in standard notation)
\begin{equation}
\label{con1}
n(x,y)\sqrt{dx^2+dy^2},
\end{equation}
i.e., the paths making
\begin{equation}
\label{con2}
\int n(x,y)\sqrt{\left(\frac{dx}{ds}\right)^2+\left(\frac{dy}{ds}\right)^2}\,ds
\end{equation}
stationary, are nicknamed {\it rays}.
The simplest realization of this Jacobi principle consists in
identifying $n^2$ with the Riemann metric tensor $g_{ij}$,
{\it whenever this identification is admissible} (see below).
This identification is obviously possible in the case of a
uniform non--absorbing medium: in this case we simply obtain a physical
realization of Euclidean geometry. However, this is also certainly possible
in the case of a refractive index of the form $n(x,y)=1/y$ $(y>0)$,
where $y$ denotes the coordinate of the vertical axis in an appropriate
reference frame. In this case we are led to the Lobacevskian metric
$ds=\sqrt{dx^2+dy^2}/y$, and the rays are the geodesics in a hyperbolic
half--plane (Poincar\'e half--plane), i.e., Euclidean half--circles with centers
on the $x$--axis (horizontal axis) and Euclidean straight lines normal
to the $x$--axis lying in the strip $0<y\leqslant 1$ of the Poincar\'e
half--plane \cite{Mishchenko}. Furthermore, in this case, the flow of geodesics
(i.e., of ray trajectories), as well as in the Euclidean case,
can be described in a rather simple and direct form, since in all spaces
where the curvature is negative or null, there are no conjugate points
at which two or more geodesics meet. But this is not the case if we consider
the geodesics on the surface of a spherical ball. In this manifold
the geodesics start spreading apart from one point, but then they begin crowding closer together
and meet, and the antipodal points of the unit sphere are conjugate.
We are then forced to past segments of geodesics together if we want to obtain
a uniform eikonal approximation. This is precisely what we have done
in Subsection \ref{subse:uniformization} (Proposition \ref{pro:2}) and in Subsection \ref{subsubse:eikonal}.

A wide extension of the GTD approach could be regarded as a sort
of ``{\it Equivalence Principle}'', which puts on equal footing
the following procedures: either the trajectories can be seen
as straight or curved lines in a Euclidean space or the trajectories
can be regarded as geodesics in a space whose metric and topological
properties are those induced by the refractive index (or, equivalently,
by the potential). However, this {\it Equivalence Principle} must be
interpreted with great caution. In fact, as we mentioned above, the identification
of $n^2$ with the Riemann metric tensor {\it is not always admissible}.
In fact, the form $g_{ij}dx^i dx^j$ must be symmetric and positive definite,
and this poses a strict restriction. Consider, for instance, a refractive index
(or a potential) of the following form: $n^2=1-V/E$ ($E<V$), where $E$ is the
energy of the incoming particle, and $V$ is the height of the potential,
as in the case of the tunnel effect (see Subsection \ref{subse:orbiting}).
In this case the geometric interpretation of the trajectory as
a real--valued geodesic in a Riemannian manifold is lost. The only possibility
remains to extend the admissible values of the phase to imaginary
and/or complex values (giving up condition (ii) indicated at the beginning of this section)
and to speak of complex rays in the sense of Landau \cite{Landau}
(see Subsection \ref{subse:orbiting}). However, it is of interest to
note that evanescent waves and, accordingly, complex phases are
generated in the shadow of the caustic as well; this is precisely the phenomenon
studied in Subsection \ref{subse:evanescent}, and, in particular, in Subsection
\ref{subsubse:solution2} in connection with the rainbow phenomenon.

\end{document}